\documentclass[aps,pra,amssymb,amsmath,eqsecnum,nofootinbib]{revtex4}
\usepackage{graphicx}

\newtheorem{definition}{Definition}[section]
\newtheorem{theorem}{Theorem}[section]
\newtheorem{proposition}{Proposition}[section]

\newtheorem{corollary}{Corollary}[section]

\newtheorem{remark}{Remark}[section]

\newenvironment{hypothesis}{HP: \begin{center}} {\end{center}}
\newenvironment{thesis}{TH: \begin{center}} {\end{center}}
\newenvironment{proof}{\begin{center}PROOF: \end{center}} {$ \blacksquare $}
\newtheorem{example}{Example}[section]
\begin{document}
\title{Time-reversal properties in the coupling of quantum angular momenta}
\author{Gavriel Segre}
\homepage{http:\\www.gavrielsegre.com}
\begin{abstract}
 After a synoptic panorama about some still unsolved foundational problems involving
 time-reversal, we show that the \emph{double time-reversal
superselection rule} of Nonrelativistic Quantum Mechanics is
redundant.

We then analyze which, among the symmetries of the Clebsch-Gordan
coefficients, may be inferred through time-reversal
considerations.

Finally, we show how Coupling Theory allows to improve our
comprehension of the fact that, in presence of hidden symmetries,
Wigner's Theorem concerning Kramers degeneration cannot be
applied, by analyzing a set of Z uncoupled massive particles, both
in the case in which they are bosons of spin zero and in the case
in which they are fermions of spin $\frac{1}{2}$, subjected to a
Keplerian force's field and considering the coupling of the 2Z
quantum angular momenta resulting by the hidden SO(4) symmetry
owed to the fact that, apart from the rotational SO(3) symmetry
responsible of the conservation of the angular momentum, the
Laplace-Runge-Lenz vector is also conserved.
\end{abstract}
\maketitle
\newpage
\tableofcontents
\newpage
\section{Introduction: a synoptical panorama of some still unsolved foundational problems involving time-reversal and menu of this paper}
The philosophical, mathematical and physical investigations
concerning the nature of Time have been pursued by the human kind
for almost three thousand years.

\emph{What is time ?}

The most common answer to this  question given by contemporary
scientists (and particularly those with an underlying strongly
positivistic attitude) is that:
\begin{enumerate}
    \item for Science time is the physical observable
    operationally defined by the clock.
    \item every other speculation is a (not particularly serious)  metaphysical intellectual
    masturbation not belonging to Science.
\end{enumerate}

Let us observe that, instead of definitely closing the issue, such
an answer opens a Pandora's box simply translating the original
question into the new question:

\emph{What is a clock ?}

whose not-triviality is linked to a plethora of reasons whose
analysis goes beyond the purposes of this paper.

We will limit ourselves to remind that:
\begin{enumerate}
    \item despite the existence of atomic clocks that, using the principle of the Maser, realize the more precise measurements of time that our actual technology allows (cfr. for instance the $ 6^{th}$ chapter
    "Two-State Systems, Principle of the Maser" of \cite{Basdevant-Dalibart-05}), the  many theorems stating in different words the impossibility of defining
    a time operator in Nonrelativistic Quantum Mechanics (cfr. for
    instance \cite{Unruh-Wald-89} and \cite{Isham-93}, the section 12.7 "The measurement of time" and the
    section 12.8 "Time and energy complementarity" of
    \cite{Peres-95} as well as
    \cite{Sala-Mayato-Alonso-Egusquiza-02}) invalidates the
    conceptual possibility of defining a genuine quantum clock.
    \item the impossibility in General Relativity of extending globally local
    clocks (cfr. the section 2.4.4 "Meanings of time" of
    \cite{Rovelli-04}).
\end{enumerate}

\smallskip

Actually we strongly think that the philosophical issue about the
nature of time is so deeply important for Physics that we cannot
leave it in the  hands of bad philosophers  \footnote{The
different philosophies of Time can be roughly divided in two
classes, according to whether they consider Time as an entity
pertaining the Subject (as in Augoustine's or, more radically, in
Immanuel Kant's thought in which the Object is the result of the
categorical synthesis of the Subject's intellectual structure) or
the Object. It has to be  remarked, with this regard, that the
Philosophy of Science underlying the Scientific Method and
properly formalized by Karl Popper's Falsification Theory
(overcoming the unphysical feature of Rudolf Carnap's Verification
Theory consisting in that it would require an infinite number of
measurements to verify a theory) is, as to the basic
epistemological issue concerning the nature of truth, essentially
a restyling of the Scholastic viewpoint \emph{"veritas est
adaequatio rei et intellectus"}. Within such a philosophical
framework, if at a foundational level time exists (it is curious,
with this regard that the idea that at a Quantum Gravity level
time doesn't exist, considered as one of the possibilities in the
$ 6^{th} $ section "Timeless Interpretation of Quantum Gravity" of
\cite{Isham-93} and explicitly advocated by Carlo Rovelli in
\cite{Rovelli-04}, \cite{Rovelli-07} coincides with the basic
philosophical statement of John Ellis Mc Taggart's "proved"
through  very involute and baroque speculations) it is an
objective property, i.e. an entity belonging to the Object of
Knowledge. Contemporary fashionable post-modern philosophies
adhering, to some degree, to Friedrich Nietsche's nihilism and/or
to the essential point of Martin Heidegger's thought, namely that
the knowledge that the Subject has of the Object is conditioned by
a pre-knowledge determined by a temporally-dependent information's
horizon into which he is thrown, giving up the solid ground of
Scientific Reason, open the door to the more dangerous
intellectual (and ethical) irrationalisms.}.

\smallskip

Of course the answer about the nature of time will be available
only when we will know the real Laws of Nature conciliating
General Relativity and Quantum Mechanics, of which we possess
nowadays a plethora of candidates whose Popper's falsifiability
exists only as a matter of principle, most of the corresponding
phenomenology being detectable only at the Planck's scale (lengths
of the order of the Planck' length $ l_{P} := \sqrt{\frac{ \hbar G
}{ c^{3} }} \approx 10^{-33} \, cm $, energies of the order of
Planck's energy  $ E_{P} := \sqrt{ \frac{ \hbar c }{G} } \approx
10^{19} \, Gev $, times of the order of Planck's time $ t_{P} :=
\sqrt{ \frac{ \hbar G  }{ c^{5} }   } \approx 10^{-44} \, s$ )
that is enormously far from our actual possibility of experimental
investigation (see for instance \cite{Kiefer-07}).

\bigskip

In this paper we will remain at a lower level taking into account
only  Non Relativistic  Quantum Mechanics that is a very good
approximation of the true Laws of Nature for enough weak
gravitational fields and for objects moving at velocities enough
smaller than the velocity of light.

At such a level, only two notions of time, among the many
mentioned by  Rovelli, will be pertinent to our analysis: the
Newtonian absolute time and the thermodynamical time.

These two notions conflict since Nonrelativistic Classical
Mechanics is symmetric under reversal of the Newtonian absolute
time, while the Second Principle states that Thermodynamics is not
invariant under the reversal of the Newtonian absolute time:

given a system of N particles leaving in the physical
three-dimensional euclidean space $( \mathbb{R}^{3} , \delta :=
\delta_{\mu \nu} dx^{\mu} \otimes dx^{\nu} )$  and hence having
hamiltonian $ H : \mathbb{R}^{6 N} \mapsto \mathbb{R} $:
\begin{equation}
    H (  \vec{q}_{1} , \vec{p}_{1} , \cdots , \vec{q}_{N} , \vec{p}_{N}) \; :=
    \; \sum_{i=1}^{N} \frac{ |p_{i}|_{\delta}^{2} }{2 m_{i}} \, +
    \, \sum_{i<j=1}^{N} V_{interaction}( \vec{q}_{i} , \vec{q}_{j} ) \, + \,
    \sum_{i=1}^{N} V_{external} ( \vec{q}_{i} )
\end{equation}
\footnote{where of course, given a vector $ \vec{v} $, we have
used the notation $ | v |_{\delta}^{2} := v^{\mu} \delta_{\mu \nu}
v^{\nu} = \vec{v}^{2} $ to emphasize its dependence from the
underlying euclidean geometry.} and defined the time-reversal
operator as the map:
\begin{equation}
    t \; \stackrel{T}{\mapsto} \; t' := -t
\end{equation}
it follows that:
\begin{equation}
    ( \vec{q}_{1} , \vec{p}_{1} , \cdots , \vec{q}_{N} ,
    \vec{p}_{N} ) (t) \stackrel{T}{\mapsto} ( \vec{q}_{1} , - \vec{p}_{1} , \cdots , \vec{q}_{N} ,
    -\vec{p}_{N} ) (t)
\end{equation}
and hence the Hamilton's equations:
\begin{equation} \label{eq:Hamilton's equations}
 \left\{%
\begin{array}{ll}
     \dot{\vec{p}}_{i} \; = \; - \partial_{ \vec{q}_{i}} H \\
    \dot{\vec{q}}_{i} \; = \; \partial_{\vec{p}_{i}} H \\
\end{array}%
\right.
\end{equation}
are T-invariant.

As a consequence, if $ c : [ t_{initial} , t_{final} ] \mapsto
\mathbb{R}^{6 N} $ is the solution of the Cauchy problem
consisting of equations \ref{eq:Hamilton's equations} and the
initial-conditions:
\begin{equation}
    \left\{%
\begin{array}{ll}
    \vec{q}_{i} ( t_{initial} ) = \vec{q}_{i}^{initial} \in \mathbb{R}^{3} \\
   \vec{p}_{i} ( t_{initial} ) = \vec{p}_{i}^{initial} \in \mathbb{R}^{3}  \\
\end{array}%
\right.
\end{equation}
and introduced:
\begin{equation}
    \left\{%
\begin{array}{ll}
    \vec{q}_{i}^{final} \; := \;  \vec{q}_{i} ( t_{final} )  \\
     \vec{p}_{i}^{final} \; := \; \vec{p}_{i} ( t_{final} )  \\
\end{array}%
\right.
\end{equation}
then $ c^{-1} ( t ) := c( t_{final} - t ) $ will be the solution
of the Cauchy problem consisting of equations \ref{eq:Hamilton's
equations} with the initial conditions:
\begin{equation}
 \left(%
\begin{array}{c}
   \vec{q}_{i} ( t_{initial} ) \\
    \vec{p}_{i} ( t_{initial} )\\
\end{array}%
\right) \; = \; T \left(%
\begin{array}{c}
 \vec{q}_{i}^{final}  \\
   \vec{p}_{i}^{final} \\
\end{array}%
\right) \; = \; \left(%
\begin{array}{c}
 \vec{q}_{i}^{final}  \\
 - \vec{p}_{i}^{final} \\
\end{array}%
\right)
\end{equation}

Contrary the Second Law of Thermodynamics holding for any isolated
thermodynamical system:
\begin{equation}
    \frac{ d S_{thermodynamic} }{ d t} \; \geq \; 0
\end{equation}
is clearly not T-invariant.

\smallskip

 It is common opinion in the Physics' scientific community (see
for instance \cite{Goldstein-01}, \cite{Lebowitz-94},
\cite{Lebowitz-07}) that Ludwig Boltzmann solved the (apparent)
paradox  through his invention of Statistical Mechanics, by
showing that a probabilistic description of a coarse-grained
representation of the reversible \footnote{Let us remark that the
term "reversible" is often used in the literature, particularly in
that devoted to "Irreversible Quantum Dynamics",  oscillating
alternatively between the following three different meanings:
\begin{description}
    \item[first meaning:] reversibility as invariance under
    time-reversal.
    \item[second meaning:] a dynamics is called reversible if it
    doesn't comport an increase of the thermodynamical entropy of the
Universe.
    \item[third meaning:] a map is called reversible if it is
    injective.
\end{description}
In this paper we will always use the term "reversible" with the
first meaning, while, when we want to refer to the second meaning,
we will use the locution "thermodynamical reversibility".}
microscopic dynamics (on which some suitable ergodicity or
chaoticity condition is eventually added \cite{Gallavotti-99a}) of
an high number of microscopic objects may generate the
phenomenological irreversibility observed macroscopically.

Indeed the expression for the thermodynamical entropy obtained
using the microcanonical ensemble:
\begin{equation}
    S( E , N , V ) \; := \; k_{Boltzmann} \,  \log ( \frac{ \int_{\Lambda} \prod_{i=1}^{N} d \vec{q}_{i} d \vec{p}_{i} \delta ( H - E )}{ N !
    })
\end{equation}
(where of course $ \int_{\Lambda} \prod_{i=1}^{N} d \vec{q}_{i} d
\vec{p}_{i} \; = V $) \footnote{beside the factor N! whose
theoretical justification can be given only considering the
classical limit of Quantum Statistical Mechanics.} summarizes the
basic Boltzmann's idea according to which:
\begin{enumerate}
    \item the thermodynamical entropy is
directly proportional to the logarithm of the number of
microstates corresponding to the given thermodynamical macrostate.
    \item the Second Law of Thermodynamical expresses the fact
    that an isolated system evolves from ordered macrostates
    (i.e. macrostates to which correspond few microstates) to
    disordered macrostates  (i.e. macrostates to which correspond many
    microstates) \footnote{It is important, with this regard, to stress that the Second Principle assumes the status of an exact law only when the thermodynamical limit $ N \rightarrow + \infty $ is performed and hence \emph{large deviations} become impossible with
    certainty.}.
\end{enumerate}

Anyway the objections that Josef Loschmidt, Henri Poincar\'{e} and
Ernst Zermelo (see for instance the $2^{th}$ chapter "Time
Reversal in Classical Mechanics" of \cite{Sachs-87}, the $ 5^{th}
$ chapter "Time irreversibility and the H-theorem" of
\cite{Cercignani-98} and \cite{Uffink-07}) moved to Boltzmann's
viewpoint are, in our modest opinion, still disturbing:
\begin{enumerate}
    \item the argument according to which the operation of acting through the
    time-reversal operator could be only  performed by a Maxwell's
    demon acting on single particles and is hence irrelevant for a
    statistical approach to the system doesn't take into account the
    following basic fact: that in many experimental situations we
    are nowadays technologically able to construct Maxwell'demons
    acting, according to some programmed algorithm, on individual
    particles of a many-particle system and that, in this case,
    the \emph{probabilistic approach to information} has to be
    considered jointly with the \emph{algorithmic approach to
    information} \footnote{It is with this regard extremely significant that the relevance of the non-probabilistic algorithmic approach to Information Theory was
    discovered by the same Andrei Nikolaevic Kolmogorov \cite{Kolmogorov-65} who is also the father of the nowadays generally accepted axiomatic measure-theoretic foundation of (Classical) Probability Theory
    \cite{Kolmogorov-56}.} by taking into account the contribution of the \emph{algorithmic
    information} of the demon to the thermodynamical entropy (see
    the $ 3^{th} $ section "Perpetual motion" of the $ 6^{th} $ chapter "Statistical Entropy" of  \cite{Penrose-05}, \cite{Landauer-90a}, \cite{Bennett-90a},
    \cite{Bennett-90b}, the $ 8^{th} $ chapter "Physics, Information and Computation" of \cite{Li-Vitanyi-97} and the $ 5^{th} $ chapter "Reversible
    Computation and the Thermodynamics of Computing" of
    \cite{Feynman-96}).
    \item Poincare's Recurrence Theorem (see for instance the Lecture I
"Measurable Transformations, Invariant Measures, Ergodic Theorems"
of \cite{Sinai-94}) states that the microscopic dynamics will
return arbitrarily close to the initial conditions.

In every course of Statistical Mechanics (see for instance the $
10^{th} $ chapter "Irreversibility and the Approach to
Equilibrium" of \cite{Schwabl-06}) it is taught that this doesn't
conflict with the thermodynamical irreversibility since an
explicit computation of the \emph{recurrence's times} shows that
they are always enormously big (of many orders of magnitude
greater than the age of the Universe) and is, in particular, of
many orders of magnitude greater than the duration of the time
interval required to reach the thermodynamical equilibrium.

 Such an answer is not, in our modest opinion, conceptually satisfying from a foundational viewpoint,
since it implies that the existence of the thermodynamical
time-asymmetry (i.e., adopting a commonly used locution, the
thermodynamical \emph{arrow of time}) would be only a transient
phenomenon:

obviously, from a mathematical viewpoint, the density of any time
interval $ [ 0 , T_{recurrence} ] $ is zero also if $
T_{recurrence} $ is, to use of the words of \cite{Gallavotti-99a},
\emph{"beyond eternity"} (i.e. of many orders of magnitude greater
than the age of the Universe):
\begin{equation}
    \lim_{t \rightarrow + \infty } \frac{\mu_{Lebesgue} ( [ 0 , T_{recurrence} ] ) }{ \mu_{Lebesgue} ( [ 0 ,t
    ])
    } \; = \; \lim_{t \rightarrow + \infty } \frac{ T_{recurrence}
    }{t} \; = \; 0 \; \; \forall T_{recurrence} \in ( 0 , + \infty
    )
\end{equation}

But we are too much convinced of the validity and importance of
the Second Principle of Thermodynamics to accept a viewpoint
according to which the existence of the \emph{thermodynamical
arrow} would be only a transient phenomenon \footnote{Such a
conviction is strengthened  taking into account, for a moment,
also General Relativity and incorporating Black-holes'
Thermodynamics (see the $ 12^{th} $ chapter "Black holes" of
\cite{Wald-84}) in the analysis.}.
\end{enumerate}

\bigskip

In the framework of Nonrelativistic Quantum Mechanics the
situation is even more delicate.

According to the orthodox Copenhagen interpretation there exist in
it two different dynamical processes:
\begin{enumerate}
    \item the unitary evolution
governing the dynamics of any \emph{closed} system (that following
Roger Penrose's terminology, see for instance \cite{Penrose-00},
we will denote as the \emph{U-process}) and ruled by the
Schr\"{o}dinger's equation.
    \item the reduction process (that
following again Penrose's terminology we will call the
\emph{R-process}) occurring when a measurement is performed on a
quantum system.
\end{enumerate}
While no doubt exists as to the time-reversal symmetry of the
\emph{U-process}, a great debate exists as to the time-reversal
symmetry of the \emph{R-process}.

If the \emph{R-process} occurring when a system $ S $ is subjected
to a measurement could be derived as a particular case of the
\emph{U-process} for the system $ S + D $ (where D is the
experimental device involved in such a measurement) then one could
immediately infer that the \emph{R-process} has time-reversal
symmetry too.

The impossibility of such a task (first attempted by John Von
Neumann in the $ 6^{th} $ chapter "The Measuring Process" of
\cite{Von-Neumann-83}) constitutes the so called \emph{Measurement
Problem of Quantum Mechanics}.

One of the more popular proposed strategies to circumvent such a
problem consists in assuming that during the measurement the
composite system S+D is weakly coupled to the environment E and is
hence a \emph{weakly open} system (defined as an \emph{open}
system for which the \emph{weak coupling limit}, for whose
mathematical definition we demand to the $ 7^{th} $ chapter "Open
Systems" of \cite{Ingarden-Kossakowski-Ohya-97}, to the $ 3^{th} $
chapter "Evolution of an Open System" of \cite{Holevo-01} or to
the $ 4^{th} $ chapter "Reversibility, Irreversibility and
Macroscopic Causality" of \cite{Sewell-02}, holds).

Since the fact that the dynamics of a \emph{closed} system is
ruled by the \emph{U-process} implies that the dynamics of a
\emph{weakly open} system is described (in the Heisenberg picture)
by a \emph{completely positive unity-preserving} dynamical
semigroup, it is then shown that such a dynamics induces a sort of
\emph{superselection rule} (see \cite{Wick-Wightman-Wigner-52},
\cite{Wightman-95},
 \cite{Zeh-03}, \cite{Giulini-03}) that cancels any superposition between
different eigenstates of the device operator furnishing the result
of the measurement.

Taking into account the Fundamental Theorem of Noncommutative
Probability \footnote{stating that the \emph{category} having as
objects the classical probability spaces and as \emph{morphisms}
the \emph{endomorphisms} (\emph{automorphisms}) of such spaces is
equivalent to the category having as objects the \emph{abelian
algebraic probability spaces} and as morphisms the endomorphisms
(\emph{automorphisms}) of such spaces; see
\cite{Kadison-Ringrose-97a},  \cite{Kadison-Ringrose-97b},
\cite{Meyer-95}, \cite{Cuculescu-Oprea-94}.} one can describe the
quantum dynamics of \emph{weakly open} systems (i.e. the Theory of
Dynamical Semigroups) in the mathematical very elegant way
available in the mentioned literature.

 Anyway nor such a mathematical restyling neither the fact of taking into account
also non-projective measurements allows to bypass the fact that,
using the words of John Bell \cite{Bell-93}, such an explanation
(called the \emph{decoherence}'s solution or the\emph{ environment
induced superselection rules'} solution) works \emph{"for all
practical purposes"} (locution usually shortened as \emph{FAPP})
but is inconsistent from a logical point of view.

In fact:
\begin{enumerate}
    \item modern technology allows to manage single particles (for instance a single photon may be obtained by attenuating the coherent output $  | z > := \exp ( - \frac{| z |^{2}}{2} ) \sum_{n=0}^{\infty} \frac{ z^{n} }{ \sqrt{n!} } | n > $ of a laser \footnote{where of course $ |  n > $ is an eigentstate of the number operator $ n:= a^{\dagger} a $ , with $ [ a , a^{\dagger}] = 1 $, of a fixed mode of the electromagnetic field.} so that $ |z| < r_{threshold} $ or by  some other more efficient photodetector; see the section 7.4 "Optical photon quantum computer" of \cite{Nielsen-Chuang-00}; for the related intriguing issue concerning
    the limit existing in acquiring information about the wave-function of a single system see \cite{Alter-Yamamoto-01}; let us stress, with this regard that the impossibility
    of the determining the wave function of a single quantum system cannot be used to infer that such a concept is not mathematically  and physically well-defined), with the
    consequence that the trick of claiming that the state of a finite quantum
    system corresponds only to an ensemble of particles and is hence specified completely by a density operator (and not by a ray) on
an Hilbert space is trivially experimentally
    falsified; all the more so such a falsification applies to interpretations according to which the state of a finite
    system is described by a ray on an Hilbert space, but corresponds only to an ensemble of particles   \footnote{Owing to the mentioned Fundamental Theorem of Noncommutative Probability, the
    statistical structure of a quantum system is indeed specified by a noncommutative probability space, i.e. a couple $ ( A , \omega
    )$ where A is a noncommutative $ W^{\star}$-algebra while $ \omega $ is a
    state on it (i.e. $ \omega \in \mathcal{S}(A)$) and hence it is ruled by a statistical structure different from the classical one. This fact, contrary to what it is sometimes claimed, doesn't anyway  solve
    the Measurement Problem of Quantum Mechanics  as we will now briefly show. Let us observe first of all that the assumption that such a quantum system is
    finite implies that the factor decomposition of A contains
    only factors of discrete type (i.e. of type  $I_{n} \, n \in
    \mathbb{N}_{+} \cup \{ + \infty \}$), that there exists a separable
    Hilbert space $ \mathcal{H} $ such that A is nothing but the algebra of all the bounded operators over $ \mathcal{H} $ (i.e. $ A =
    \mathcal{B} ( \mathcal{H} )) $, and that the state $ \omega $ is
    normal and hence there exists a density operator $ \rho $ such
    that $ \omega( a) \; = \; Tr ( \rho \, a) \; \; \forall a \in A
    $. In the Heisenberg's picture of the motion the U-process is
    described by a strongly continuous group of inner automorphisms
    of A. Let us now observe that in the particular case in which the system is a single
    particle the removal of any epistemic ignorance about it
    implies that the system must be described by in a extremal state $
    \omega \in \Xi (A) $, i.e.
    by a state corresponding to a pure density matrix  $ \rho = | \psi > < \psi |
    $ (for a suitable  $ | \psi > \in \mathcal{H} $)  and the Measurement Problem appears again in its usual form.}.
    \item there is no reason to assume that, as a matter of principle, it is impossible to
    isolate the system S+D and hence to treat it as a closed
    system.
    \item such an explanation simply translates the problem of obtaining the
    \emph{R-process} for S as an \emph{U-process} for S+D into  the problem of obtaining it from the \emph{U-process} for the
    composite system S + D + E \footnote{The same can  be said as to the iteration of such a procedure in which one introduces a sequence of devices $ \{ D_{n} \}_{n \in \mathbb{N}_{+}} $ and, given $ n \in \mathbb{N}_{+}$,  one attempts to derive
    the \emph{R-process} for $ S + D_{1} ,  \cdots + D_{n-1} $  as an
    U-process for $ S + D_{1} ,  \cdots + D_{n} $. At every finite $ n \in \mathbb{N}_{+} $ level of this chain, called  \emph{the Von Neumann's chain}, one faces a conceptual problem identical to the original one. If contrary one attempts to perform the limit $ n \rightarrow + \infty $ the
    emerging peculiarities  of the quantum theory of infinite systems (such as the existence of unitarily inequivalent representations of the observables' algebra; see for instance \cite{Strocchi-85}, \cite{Strocchi-05b}, \cite{Strocchi-05a}) may contribute to hide the problem under a mountain of mathematical sophistication
    but not to solve it.}.
\end{enumerate}

Decoherence is actually a dramatically true phenomenon responsible
for the appearance of a  macroscopical classical behavior from an
underlying microscopic Quantum Theory (and a hated enemy fought
through quantum error-correcting codes with success only for a few
qubits by those involved in the physical implementation of
quantum-computation; see for instance the $ 10^{th} $ chapter
"Quantum error-correction" of \cite{Nielsen-Chuang-00}) but is
not, in our modest opinion, a conceptually consistent solution to
the \emph{Measurement Problem of Quantum Mechanics}.

What  is relevant to our present purposes is anyway simply to
stress that, whichever viewpoint is assumed as to the
interpretational problems of Quantum Mechanics and despite the
opposite claim so often uncritically assumed in the literature,
there are reasons to suppose the also the \emph{R-process} may
occur in a time-reversal invariant way (see for instance
\cite{Aharonov-Bergmann-Lebowitz-83}), with the net effect that
the conflict between the microscopic reversibility and the
macroscopic thermodynamical irreversibility appears again, with
the considerations previously exposed as to Classical Statistical
Mechanics easily extendible to the quantum case.

The reasons to assume that probabilistic considerations are not
sufficient to explain the existence of the arrow of time are, in
that context, even more striking, going from the interplay between
Lorentzian Geometry and Thermodynamics showed by the appearance of
Kubo-Martin-Schwinger thermal states in Quantum Field Theories on
spacetimes presenting a bifurcate Killing horizon \cite{Wald-94}
to arguments of Quantum Cosmology going far beyond our competence
(see for instance \cite{Gell-Mann-Hartle-94}, \cite{Hawking-94},
\cite{Halliwell-94}, the $ 5^{th} $ chapter "The Time Arrow of
spacetime" and the $ 6^{th} $ chapter "The Time Arrow in Quantum
Cosmology" of \cite{Zeh-07}).

\bigskip

After the presented synoptic view about the time-reversal
foundational issues  let us at last present the content, strongly
more modest, of this paper:

to discuss some time-reversal properties in the coupling of
quantum angular momenta that might contribute to shed some little
light on them.

More specifically, the menu of this paper is the following:
\begin{itemize}
    \item in the section \ref{sec:Coupling and recoupling of quantum angular momenta} we briefly review
    some basic facts concerning the Coupling Theory and the
    Recoupling Theory of quantum angular momenta.
    \item in the section \ref{sec:Considerations about the double
time-reversal superselection rule in Nonrelativistic Quantum
Mechanics} we show that that the \emph{double time-reversal
superselection rule} of Nonrelativistic Quantum Mechanics is
redundant.

\item in the section \ref{sec:Symmetries of the Clebsch-Gordan
coefficients deducible from time-reversal considerations} we
analyze which, among the symmetries of the Clebsch-Gordan
coefficients, may be inferred through time-reversal
considerations.

\item in the section \ref{sec:Kramers degenerations, Hidden
Symmetries and Coupling Theory} we show how Coupling Theory allows
to improve our comprehension of the fact that, in presence of
hidden symmetries, Wigner's Theorem concerning Kramers
degeneration cannot be applied, by analyzing a set of Z uncoupled
massive particles, both in the case in which they are bosons of
spin zero and in the case in which they are fermions of spin
$\frac{1}{2}$, subjected to a Keplerian force's field and
considering the coupling of the 2Z quantum angular momenta
resulting by the hidden SO(4) symmetry pertaining to the fact
that, apart from the rotational SO(3) symmetry, the
Laplace-Runge-Lenz vector is conserved.

\end{itemize}

We would like to remark that, though not being a paper about
Quantum Gravity, this paper could be of some interest for those
engaged in such a research field owing to the astonishing link
existing between General Relativity and the Theory of Quantum
Angular Momentum discovered by Giorgio Ponzano and Tullio Regge in
1968 (i.e that for large values of the quantum number j the Wigner
6j symbol approximates the Einstein-Hilbert action; see
\cite{Ponzano-Regge-68}, the $ 9^{th} $ topic "Physical
Interpretation, and Asymptotic (Classical) Limits of the Angular
Momentum Functions" of the $ 5^{th} $ chapter "Special Topics" of
\cite{Biedenharn-Louck-81} and the section 7.4.3 "Construction in
terms of 6j symbols" of \cite{Ambjorn-Durhuus-Jonsson-97}) and its
deduction in the framework of Loop Quantum Gravity where the
physical states of the quantum gravitational field are
spin-networks (see the $ 6^{th} $ chapter "Quantum space", the $
9^{th}$ chapter "Quantum spacetime: spinfoams" and the appendix A
"Groups and recoupling theory" of \cite{Rovelli-04}).

\newpage
\section{Coupling and recoupling of quantum angular momenta} \label{sec:Coupling and recoupling of quantum angular momenta}
\begin{figure}
  \includegraphics[scale=.5]{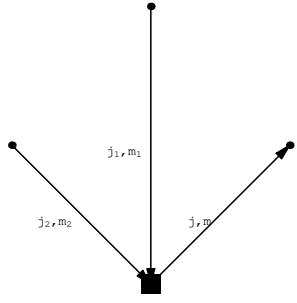}\\
  \caption{The Clebsch-Gordan diagram}\label{fig:Clebsch-Gordan diagram}
\end{figure}

Let us recall  \cite{Edmonds-60},
\cite{Varshalovich-Moskalev-Khersonskii-88}, \cite{Thomson-2004}
that the coupling of two quantum angular momenta $ \vec{J}_{1},
\vec{J}_{2} $ may be represented graphically by a labelled diagram
with one vertex (corresponding to the Clebsch-Gordan coefficient $
C_{j_{1}m_{1}j_{2}m_{2}}^{jm} $) having two entering edges (one
corresponding to the couple of quantum numbers $ ( j_{1} , m_{1} )
$ and the other corresponding to the couple of quantum numbers $ (
j_{2} , m_{2}))$ and one exiting edge (corresponding to the couple
of quantum numbers $ (j, m )$ ) as represented in the figure
\ref{fig:Clebsch-Gordan diagram}.

We will call such a graph the Clebsch-Gordan diagram.

Using the new version of the standard package \emph{Combinatorica}
for \emph{Mathematica 5} \cite{Wolfram-03} described in
\cite{Pemmaraju-Skiena-03}, the Clebsch-Gordan diagram may be
trivially implemented by the following Mathematica expression
\verb"cgdiagram[x,y]" where \verb"x" and \verb"y"  have to be the
list of the labels of, respectively, the $ 1^{th} $ and the $
2^{th} $ coupled angular momentum:
\begin{verbatim}
<< DiscreteMath`Combinatorica`

cgdiagram[x_, y_] :=
  SetGraphOptions[
    FromOrderedPairs[{{1, 3}, {2, 3}, {3, 4}}], {3,
      VertexStyle -> Box[Large]},
    EdgeLabel -> {"j" <> ToString[x] <> "m" <> ToString[x],
        "j" <> ToString[y] <> "m" <> ToString[y],
        "j" <> ToString[Join[x, y]] <> "m" <> ToString[Join[x, y]]}]
\end{verbatim}

Demanding to the mentioned literature for the many existing
explicit expressions for the Clebsch-Gordan coefficient and its
meaning as the probability  amplitude $
C_{j_{1}m_{1}j_{2}m_{2}}^{jm} := < j_{1} , j_{2} , m_{1} , m_{2}
|j_{1} , j_{2} , j , m > $, the basic thing we will need to know
about it is that it is always real and that:
\begin{proposition} \label{prop:basic property of the Clebsch-Gordan coefficient}
\end{proposition}
\begin{multline}
  C_{j_{1}m_{1}j_{2}m_{2}}^{jm} \, \neq 0 \, \; \Rightarrow \; j
  \in \frac{\mathbb{N}}{2}  \cap [ | j_{1} -
  j_{2}| , j_{1} + j_{2} ] \, \wedge \, m = m_{1} + m_{2} \\
  \forall m_{1} \in \{ - j_{1}, \cdots , j_{1} \},  \forall m_{2} \in \{ - j_{2}, \cdots , j_{2} \} ,  \forall m \in \{ - j, \cdots , j \}, \forall  j_{1},j_{2},j \in \frac{\mathbb{N}}{2}
\end{multline}
where, given $ a , b \in \mathbb{R} : b-a \in \mathbb{N} $, we
have adopted the notation:
\begin{equation}
    \{ a , \cdots , b \} \; := \; \{ a + k \, , \, k \in
    \mathbb{N} \cap [0 , b -a ] \}
\end{equation}

Let us now observe that the proposition \ref{prop:basic property
of the Clebsch-Gordan coefficient} implies that:
\begin{corollary} \label{cor:allowed j are equally spaced of a unit step}
\end{corollary}
\begin{multline}
  C_{j_{1}m_{1}j_{2}m_{2}}^{jm} \, \neq 0 \, \; \Rightarrow \; j
  \in \{ |j_{1}-j_{2}| , \cdots , j_{1} + j_{2} \}  \\
  \forall m_{1} \in \{ - j_{1}, \cdots , j_{1} \},  \forall m_{2} \in \{ - j_{2}, \cdots , j_{2} \} ,  \forall m \in \{ - j, \cdots , j \}, \forall  j_{1},j_{2},j \in \frac{\mathbb{N}}{2}
\end{multline}
\begin{proof}
Let us consider the different cases:
\begin{enumerate}
    \item in the case $ j_{1},j_{2} \in \mathbb{N} $ then obviously
   $ | j_{1} - j_{2} | \in \mathbb{N} $. Since $ m_{1} \in \{ - j_{1} ,
   \cdots , j_{1} \} $ and  $ m_{2} \in \{ - j_{2} ,
   \cdots , j_{2} \} $ it follows that $ m_{1} , m_{2} \in \mathbb{N}
   $ and hence, obviously, $ m_{1} + m_{2} \in \mathbb{N} $.

   Let us consider a  $ j \in \frac{\mathbb{N}}{2}  \cap [ | j_{1} -
  j_{2}| , j_{1} + j_{2} ] $ such that $ j -  | j_{1} - j_{2} |
  \notin \mathbb{N} $ that is, following the notation of section \ref{sec:Some useful algebraic
  properties}, $ j \in \mathbb{H} $. Since $ m \in \{ - j ,
   \cdots , j \} $, the proposition \ref{prop:algebra of the integers union the
   half-integral} implies that $ m \in \mathbb{H} $ with the consequence that $ m
   \neq m_{1} + m_{2} $ and hence $ C_{j_{1}m_{1}j_{2}m_{2}}^{jm} \,
   = \, 0 $.

    \item in the case $ j_{1},j_{2} \in \mathbb{H} $ then, by the proposition \ref{prop:algebra of the integers union the
   half-integral} $ | j_{1} - j_{2} | \in \mathbb{N} $. Since $ m_{1} \in \{ - j_{1} ,
   \cdots , j_{1} \} $ and  $ m_{2} \in \{ - j_{2} ,
   \cdots , j_{2} \} $ it follows that $ m_{1} , m_{2} \in
   \mathbb{N}$ and hence, obviously, $ m_{1} + m_{2} \in \mathbb{N} $.

    Let us consider a  $ j \in \frac{\mathbb{N}}{2} \cap [ | j_{1} -
  j_{2}| , j_{1} + j_{2} ] $ such that $ j -  | j_{1} - j_{2} |
  \notin \mathbb{N} $ that is $ j \in \mathbb{H} $. Since $ m \in \{ - j ,
   \cdots , j \} $, the proposition \ref{prop:algebra of the integers union the
   half-integral} implies that $ m \in \mathbb{H} $  with the consequence that $ m
   \neq m_{1} + m_{2} $ and hence $ C_{j_{1}m_{1}j_{2}m_{2}}^{jm} \,
   = \, 0 $.

   \item in the case $ j_{1} \in \mathbb{N} $ while $ j_{2} \in
   \mathbb{H}$ then, by the proposition \ref{prop:algebra of the integers union the
   half-integral} $ | j_{1} - j_{2} | \in \mathbb{H} $.  Since $ m_{1} \in \{ - j_{1} ,
   \cdots , j_{1} \} $ and  $ m_{2} \in \{ - j_{2} ,
   \cdots , j_{2} \} $, the proposition \ref{prop:algebra of the integers union the
   half-integral} implies that $ m_{1} \in \mathbb{N} $, that $
   m_{2} \in \mathbb{H} $ and hence that $ m_{1} + m_{2} \in \mathbb{H} $.

   Let us consider a  $ j \in  \frac{\mathbb{N}}{2} \cap [ | j_{1} -
  j_{2}| , j_{1} + j_{2} ] $ such that $ j -  | j_{1} - j_{2} |
  \notin \mathbb{N} $ that is $ j \in \mathbb{N} $. Since $ m \in \{ - j ,
   \cdots , j \} $, the proposition \ref{prop:algebra of the integers union the
   half-integral} implies that $ m \in \mathbb{N} $  with the consequence that $ m
   \neq m_{1} + m_{2} $ and hence $ C_{j_{1}m_{1}j_{2}m_{2}}^{jm} \,
   = \, 0 $.
\end{enumerate}
\end{proof}

\smallskip

Furthermore, since $ | C_{j_{1}m_{1}j_{2}m_{2}}^{jm} |^{2} $ is
the probability that a measurement of $ \vec{J}^{2}:=
J_{1}^{2}+J_{2}^{2} + J_{3}^{2} $ and $ J_{3} $  when the system
is in the state $ | j_{1} ,j_{2}, m_{1} , m_{2}
> $ give, respectively, the value $ j( j+1 ) $ and the value m, the
normalization of probabilities, combined with the application of
the proposition \ref{prop:basic property of the Clebsch-Gordan
coefficient} and the corollary \ref{cor:allowed j are equally
spaced of a unit step}, implies that:
\begin{proposition} \label{prop:normalization of the probability for 1 Clebsch-Gordan coefficients}
\end{proposition}
\begin{equation}
    \sum_{j= | j_{1} - j_{2} |}^{j_{1}+j_{2}} \sum_{m = - j }^{j}
    | C_{j_{1}m_{1}j_{2}m_{2}}^{jm} |^{2}  \; = \; 1
\end{equation}

\bigskip

Let us now suppose to have $ n \in \mathbb{N} : n \geq 2 $ quantum
angular momenta $ \vec{J}_{1} , \cdots , \vec{J}_{n}$.

Introduced the operators:
\begin{equation}
    \vec{J} \: := \; \sum_{i=1}^{n} \vec{J}_{i}
\end{equation}
\begin{equation}
    \vec{J}_{(1 \cdots k)} \; := \;  \sum_{i=1}^{k} \vec{J}_{i} \;
    \; k \in \{ 2 , \cdots , n-1 \}
\end{equation}
we have that each of the possible $ ( 2n- 3 ) !! $ \footnote{Let
us recall that:
\begin{equation}
    n !! \; := \; \left\{%
\begin{array}{ll}
    \prod_{ \{ k \in \mathbb{E} : k \leq n \} } k, & \hbox{if $ n \in \mathbb{E}$;} \\
    \prod_{ \{ k \in \mathbb{O} : k \leq n \} } k, & \hbox{f $ n \in \mathbb{O}$.} \\
\end{array}%
\right.
\end{equation}
has not to be confused with:
\begin{equation}
    (n!)! \; = \; \prod_{k=1}^{n!} k \, = \,  \prod_{k=1}^{\prod_{l=1}^{n} l } k
\end{equation}} \emph{coupling schemes} may be represented by a labelled
oriented tree with n entering edges (corresponding to the n
couples of quantum numbers $ (j_{1},m_{1}) , \cdots , (j_{n} ,
m_{n} ) $) and one exiting edge (corresponding to (j,m) ) with
(n-1) vertices, each with 2 entering edges and one exiting edge;
clearly all these labelled oriented trees may be obtained gluing
together in all the possible ways n-1 Clebsch-Gordan diagrams,
operation that, in the Mathematica 5 setting described above, may
be easily implemented through suitable combinations of the
Mathematica expressions \verb"Contraction", \verb"AddEdges" and
\verb"DeleteEdges" combined with the following Mathematica
expressions useful to manage the  \emph{symmetric group of order
n} $ S_{n} $, i.e. the group of all the permutations of n objects:
\begin{verbatim}
symmetrize[f_,listofarguments_]:=(1/Factorial[Length[listofarguments]])*
    Sum[Apply[f,
        Permute[listofarguments,
          UnrankPermutation[p,Length[listofarguments]]]],{p,1,
        Factorial[Length[listofarguments]]}]

antisymmetrize[f_,listofarguments_]:=(1/Factorial[Length[listofarguments]])*
    Sum[SignaturePermutation[UnrankPermutation[p,Length[listofarguments]]]*
        Apply[f,Permute[listofarguments,
            UnrankPermutation[p,Length[listofarguments]]]],{p,1,
        Factorial[Length[listofarguments]]}]

exchange[listofarguments_,couple_]:=
  Permute[listofarguments,
    FromCycles[
      Join[{{couple[[1]],couple[[2]]}},
        Table[{Complement[Range[Length[listofarguments]],couple][[i]]},{i,1,
            Length[listofarguments]-2}]]]]

operatorofexchange[f_,listofarguments_,couple_]:=
  Apply[f,exchange[listofarguments,couple]]
\end{verbatim}

\begin{figure}
  \includegraphics[scale=.5]{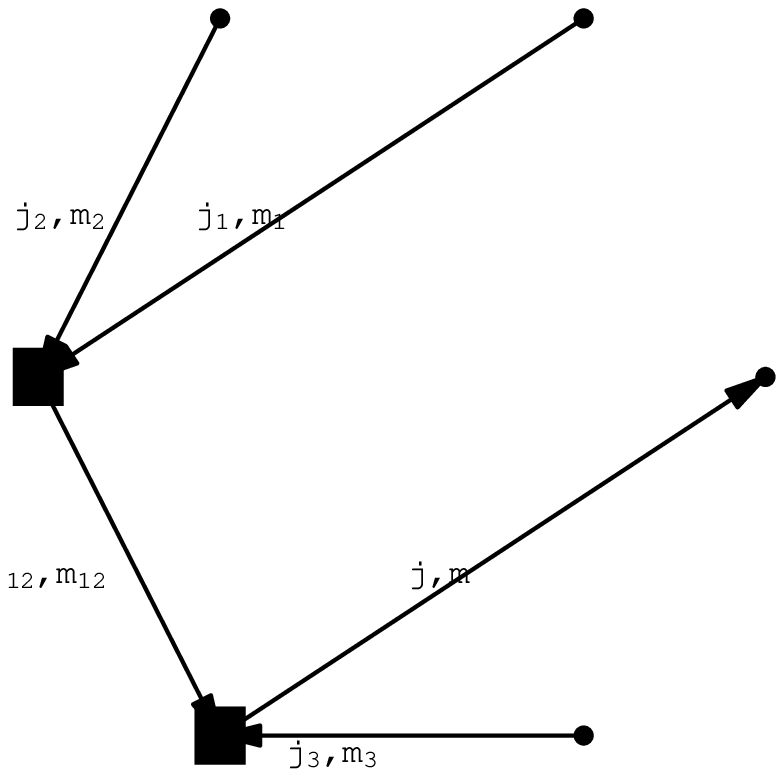}
  \includegraphics[scale=.5]{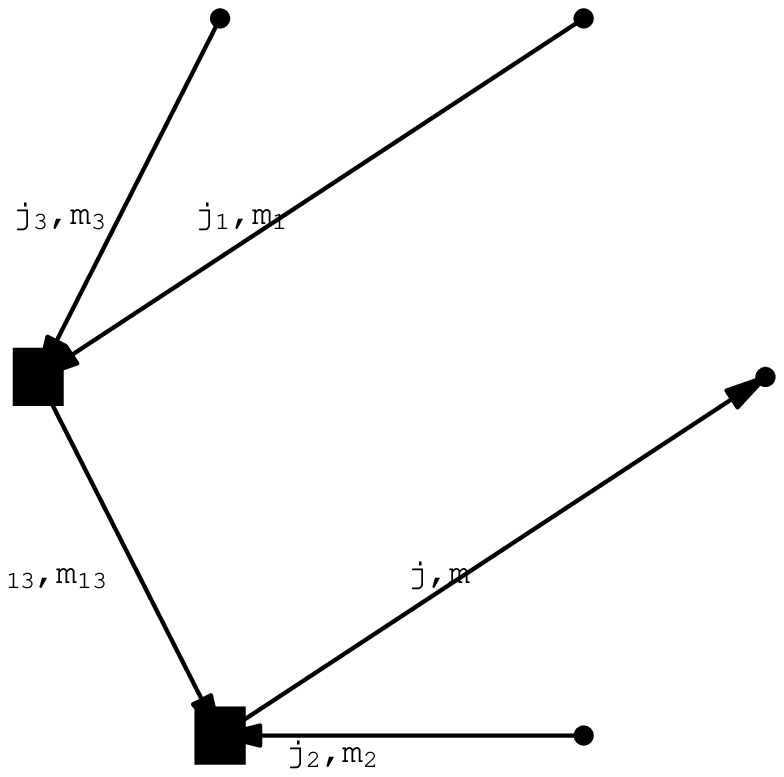}
  \includegraphics[scale=.5]{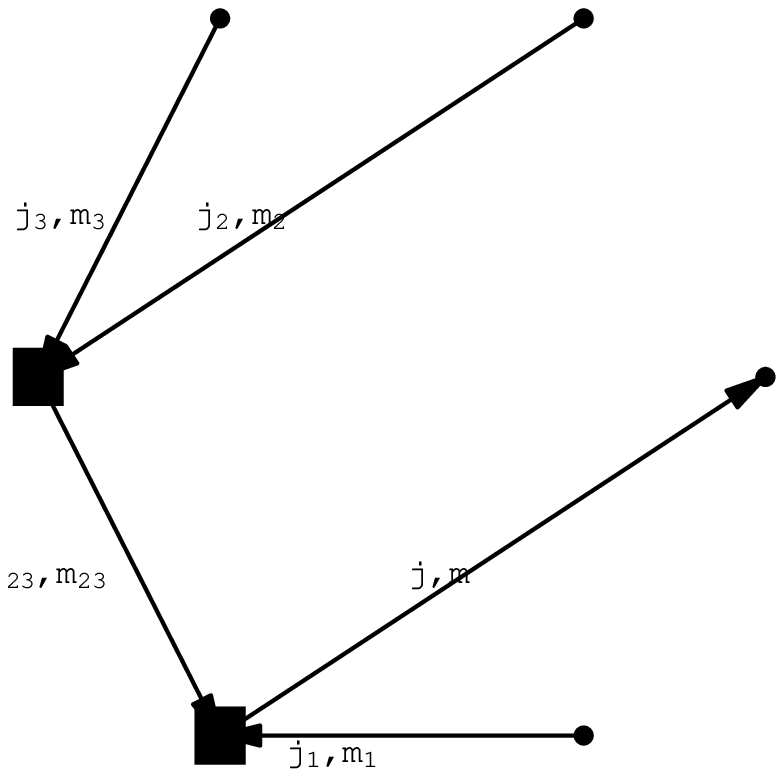} \\
  \caption{The diagrams corresponding to the three \emph{coupling schemes} of three angular momenta} \label{fig:coupling diagrams of three angular momenta}
\end{figure}

For instance the three \emph{coupling schemes} existing in the
coupling of three angular momenta are represented in the figure
\ref{fig:coupling diagrams of three angular momenta}.

The complete orthonormal bases corresponding to different
\emph{coupling schemes} are related by suitable unitary operators
whose information is completely encoded in  the \emph{Wigner
(3n)-j symbol}. The operation of passing from one \emph{coupling
scheme} to an other is called recoupling and the associated theory
is called Recoupling Theory, a theory whose representability in
diagrammatic way is deeply linked to Knot Theory (see  the
appendix A "Groups and recoupling theory" of \cite{Rovelli-04},
\cite{Kaufmann-Lins-94}  \cite{Carter-Flath-Saito-95} and the
section 2.13 "Supplement on Combinatorial Foundations" of
\cite{Louck-06} for any further information).

In the following we will consider only one of these \emph{coupling
schemes}, namely the one in which the $ \{ \vec{J}_{A}
\}_{A=1}^{n} $ are coupled sequentially:
\begin{equation}
    \vec{J}_{1} +  \vec{J}_{2} \, = \, \vec{J}_{12} \; , \;
    \vec{J}_{12} + \vec{J}_{3} \, = \, \vec{J}_{123} \; , \;
    \cdots \; , \vec{J}_{1 \cdots n-1} + \vec{J}_{n} \, = \, \vec{J}
\end{equation}

 Given $ j_{1} , \cdots , j_{n} \in
\frac{\mathbb{N}}{2} $ the eigenvectors corresponding to the two
complete set of commuting observables $ \{ \vec{J}_{1} , J_{1 z} ,
\vec{J}_{2} , J_{2 z} , \cdots , \vec{J}_{n} , J_{n z} \} $ and $
\{ \vec{J}_{1} ,  \vec{J}_{2} , \vec{J}_{12}, \vec{J}_{3} ,
\vec{J}_{123}, \cdots , \vec{J}_{n-1} , \vec{J}_{1 \cdots n-1} ,
\vec{J} , J_{z} \} $ are linked by:

\begin{proposition} \label{prop:coupling of n angular momenta}
\end{proposition}
\begin{multline}
    | j_{1}, j_{2}, j_{12} , j_{3} , j_{123} , \cdots , j_{1 \cdots
    n-1} , j_{n} , j, m > \; = \\ \sum_{m_{1}=-j_{1}}^{j_{1}}
    \sum_{m_{2}=-j_{2}}^{j_{2}} \cdots \sum_{m_{n}=-j_{n}}^{j_{n}}
      C_{j_{1 \cdots n-1}m_{1 \cdots n-1}j_{n}m_{n}}^{jm}   \cdots   C_{j_{12}m_{12}j_{3}m_{3}}^{j_{123}m_{123}}   C_{j_{1}m_{1}j_{2}m_{2}}^{j_{12}m_{12}}
      | j_{1}, m_{1} , \cdots , j_{n}, m _{n} > \\
      \forall m \in \{ - j , \cdots , j \} , \forall j \in \{| j_{1 \cdots n-1} - j_{n} | , \cdots , j_{1 \cdots n-1}
      +j_{n} \} \\
 \forall j_{1 \cdots n-1} \in \{| j_{1 \cdots n-2} - j_{n-1} | , \cdots , j_{1 \cdots n-2}
      +j_{n-1} \} , \cdots, \forall j_{123} \in \{| j_{12} - j_{3} | , \cdots , j_{12}
      +j_{3} \} , \forall j_{12} \in \{| j_{1} - j_{2} | , \cdots , j_{1}
      +j_{2} \}
\end{multline}

where we have adopted (as we will implicitly do from now on) the
notation:
\begin{equation}
    m_{12} \; := \; m_{1} + m_{2}
\end{equation}
\begin{equation}
    m_{1 \cdots k} \; := \;  m_{1 \cdots k-1} + m_{k} \; = \;
    \sum_{i=1}^{k} m_{i} \; \; \forall k \in \{ 3 , \cdots , n \}
\end{equation}

It may be useful to introduce the following:

\begin{definition} \label{def:generalized coupling coefficient}
\end{definition}
\emph{generalized coupling coefficient:}
\begin{multline}
    \mathcal{C}_{j_{1}, m_{1} , \cdots , j_{n}, m _{n}}^{ j_{1}, j_{2}, j_{12} , j_{3} , j_{123} , \cdots , j_{1 \cdots
    n-1} , j_{n} , j, m } \; := \\
     <  j_{1}, m_{1} , \cdots , j_{n},
    m_{n}  | j_{1}, j_{2}, j_{12} , j_{3} , j_{123} , \cdots , j_{1 \cdots
    n-1} , j_{n} , j, m > \; = \; C_{j_{1 \cdots n-1}m_{1 \cdots n-1}j_{n}m_{n}}^{jm}   \cdots   C_{j_{12}m_{12}j_{3}m_{3}}^{j_{123}m_{123}}   C_{j_{1}m_{1}j_{2}m_{2}}^{j_{12}m_{12}}
\end{multline}

implemented computationally by the following Mathematica
expression:
\begin{verbatim}
generalizedcouplingcoefficient[listofpedices_, listofapices_] :=
  Product[ ClebschGordan[{listofapices[[2*i - 1]] ,
        Sum[listofpedices[[2*j]], {j, 1, i}] }, {listofapices[[2*i]],
        listofpedices[[2 + 2*i]]}, {listofapices[[2*i + 1]],
        If[i == Length[listofpedices]/2 - 1, Last[ listofapices ]  ,
          Sum[listofpedices[[2*j]], {j, 1, i + 1}] ]}], {i, 1,
      Length[listofpedices]/2 - 1}]
\end{verbatim}

Let us observe that clearly the quantum number j can take the
following values:
\begin{equation}
   j \in \{ \mathcal{J}_{min} ( j_{1} , \cdots , j_{n} ) ,  \cdots ,  \mathcal{J}_{max} ( j_{1} , \cdots , j_{n} )  \}
\end{equation}
where the map $ \mathcal{J}_{max} : (\frac{\mathbb{N}}{2})^{n}
\mapsto \frac{\mathbb{N}}{2} $ is defined simply as:
\begin{equation}
  \mathcal{J}_{max} ( j_{1} , \cdots , j_{n} ) \; := \;
  \sum_{i=1}^{n} j_{i}
\end{equation}
while the map $ \mathcal{J}_{min} : ( \frac{\mathbb{N}}{2} )^{n}
\mapsto \frac{\mathbb{N}}{2} $ is defined recursively as:
\begin{equation}
\mathcal{J}_{min} ( j_{1} , \cdots , j_{n} )  \; := \;
\left\{%
\begin{array}{ll}
   | j_{1} - j_{2} |, & \hbox{if $ n = 2 $;} \\
    \min \{ | i - j_{n} |   \, , \, i \in  \{ \mathcal{J}_{min} ( j_{1} , \cdots , j_{n-1} ) , \cdots , \mathcal{J}_{max} ( j_{1} , \cdots , j_{n-1} ) \} \}, & \hbox{otherwise.} \\
\end{array}%
\right.
\end{equation}
Both these maps may be computed through the following Mathematica
5 expressions \verb"jmin[x]" and \verb"jmax[x]", where \verb"x" is
the list $ \{ j_{1} , \cdots , j_{n} \}$:
\begin{verbatim}
jmaximum[x_,n_]:=Sum[x[[i]],{i,1,n}]

jminimum[x_,n_]:=
  If[n==2, Abs[x[[1]]-x[[2]]],
    Min[Table[Abs[i-x[[n]]] ,{i,jminimum[x,n-1],jmaximum[x,n-1]}]]]

jmin[x_]:=jminimum[x,Length[x]]

jmax[x_]:=jmaximum[x, Length[x]]
\end{verbatim}

\smallskip

Taking into account all these considerations together with the
proposition \ref{prop:basic property of the Clebsch-Gordan
coefficient} it follows that:
\begin{proposition}
\end{proposition}
\begin{equation}
  \mathcal{C}_{j_{1}, m_{1} , \cdots , j_{n}, m _{n}}^{ j_{1}, j_{2}, j_{12} , j_{3} , j_{123} , \cdots , j_{1 \cdots
    n-1} , j_{n} , j, m }  \neq 0 \; \Rightarrow \;
    j \in \{ \mathcal{J}_{min}( j_{1} , \cdots , j_{n} ) , \cdots
    , \mathcal{J}_{max}( j_{1} , \cdots , j_{n} ) \} \, \wedge \, m =
    \sum_{i=1}^{n} m_{i}
\end{equation}

\smallskip

Furthermore the definition \ref{def:generalized coupling
coefficient} and the proposition \ref{prop:normalization of the
probability for 1 Clebsch-Gordan coefficients} imply that:
\begin{proposition} \label{prop:normalization of the probability for the generalized coupling coefficient}
\end{proposition}
\begin{equation}
    \sum_{j_{12}= | j_{1} - j_{2}| }^{ j_{1} + j_{2}}  \sum_{j_{123}= | j_{12} - j_{3}| }^{ j_{12} +
    j_{3}} \cdots \sum_{j = | j_{1 \cdots n-1} - j_{n} |}^{j_{1 \cdots n-1} +
    j_{n}} \sum_{m=-j }^{j} |    \mathcal{C}_{j_{1}, m_{1} , \cdots , j_{n}, m _{n}}^{ j_{1}, j_{2}, j_{12} , j_{3} , j_{123} , \cdots , j_{1 \cdots
    n-1} , j_{n} , j, m }  |^{2} \; = \; 1
\end{equation}

\newpage
\section{Considerations about the \emph{double time-reversal superselection rule} in Nonrelativistic Quantum
Mechanics} \label{sec:Considerations about the double
time-reversal superselection rule in Nonrelativistic Quantum
Mechanics}

 The notion of \emph{superselection rule} introduced by
Gian Carlo Wick , Arthur Wightman and Eugene Wigner in their
celebrated paper \cite{Wick-Wightman-Wigner-52} is nowadays a
corner stone of our comprehension of the mathematical structure of
(Special)-Relativistic Quantum Mechanics  \cite{Weinberg-95},
\cite{Streater-Wightman-00},  \cite{Haag-96} (i.e. of Quantum
Field Theory over the Minkowskian spacetime $ ( \mathbb{R}^{4} ,
\eta ) $ \footnote{where of course $ \eta := \eta_{\mu \nu}
dx^{\mu} \otimes dx^{\nu} $, $ \eta_{\mu \nu} := diag( -1,1,1,1)
$.}).

Since Nonrelativistic Quantum Mechanics may be obtained from
Quantum Field Theory over the Minkowski spacetime by taking the
limit $ c \rightarrow + \infty $, implemented
    mathematically as the contraction \cite{Barut-Racza-86} of the Lie algebra $ L[ \mathcal{P} ] $ of the isometries's group
of the Minkowski spacetime  \footnote{i.e. the Lie algebra $ L[
\mathcal{P}] $ of the Poincar\'{e} group $ \mathcal{P} $, Lie
algebra generated by the Killing vector
    fields $ P_{\mu} $ and $ M_{\mu \nu} $:
\begin{equation}
    [ P_{\mu} , P_{\nu} ] \; = \; 0
\end{equation}
\begin{equation}
    M_{\nu \mu} \; = \; - M_{\mu \nu}
\end{equation}
\begin{equation}
    [ M_{\mu \nu} , M_{\rho \sigma} ] \; = \; \eta_{\nu \rho} M_{\mu
    \sigma} + \eta_{\mu \sigma} M_{\nu \rho} - \eta_{\mu \rho } M_{\nu
    \sigma} - \eta_{\nu \sigma} M_{\mu \rho}
\end{equation}
\begin{equation}
    [ M_{\mu \nu} , P_{\sigma} ] \; = \; \eta_{\mu \sigma} P_{\nu} -
    \eta_{\nu \sigma} P_{\mu}
\end{equation}
whose contraction results in the Lie algebra of the Galilei group
admitting the basis $ P_{\mu} , K_{i} $ such that:
\begin{equation}
    M_{0i} \; =: \; c K_{i}
\end{equation}
\begin{equation}
    [ P_{\mu} , P_{\nu} ] \; = \; 0
\end{equation}
\begin{equation}
    [ K_{i} , K_{j} ] \; = \; \lim_{c \rightarrow  +
    \infty} \frac{1}{c^{2}} [ M_{0i} , M_{0j} ] \; = \; 0
\end{equation}
\begin{equation}
    [ K_{i} , P_{\mu} ] \; = \; \lim_{c \rightarrow  +
    \infty} \frac{1}{c} [ M_{0i} , P_{\mu} ] \; = \; 0
\end{equation}}, it would seem reasonable to suppose that also the status of
superselection rules in Nonrelativistic Quantum Mechanics should
be a settled issue.

Curiously the status of the \emph{double time-reversal
superselection rule} is Nonrelativistic Quantum Mechanics is not
yet clear.

The \emph{double time-reversal superselection rule} states that
the square $ T^{2} $ of the (antilinear, antiunitary)
\emph{time-reversal operator} T (about which see for instance the
$ 26^{th} $ chapter "Time Inversion" of \cite{Wigner-59}, the
Topic 1 "Fundamental Symmmetry Considerations" of the $ 5^{th} $
chapter "Special Topics" of \cite{Biedenharn-Louck-81}, the
$15^{th}$ chapter "Invariance and Conservation Theorems, Time
Reversal" of \cite{Messiah-61}, the $ 4^{th} $ chapter "Symmetries
in Quantum Mechanics" of \cite{Sakurai-94}, the $ 13^{th} $
chapter "Discrete Symmetries" of \cite{Ballentine-98}, the $
10^{th} $ chapter "Time Reversal" of \cite{Lax-01} and, last but
not least, \cite{Sachs-87}) is a \emph{superselection charge}.

Let us define as \emph{even under double time-reversal} any state
$ | \psi> $ such that $ T^{2} | \psi > =   | \psi > $.

Similarly, let us define as \emph{odd under double time-reversal}
any state $ | \psi
> $ such that $ T^{2} | \psi > =  -  | \psi > $.

The physical argument proposed by Wick, Wightman and Wigner
supporting the existence of the \emph{double time-reversal
superselection rule} is the following:

since inverting twice the arrow of time should have no physical
effect, given a state $ | E
> $ even under double time-reversal  and a state $  | O
> $ odd under double time-reversal, one should have that the states $ | E >
+ | O > $ and the state $ T^{2} ( | E > + | O >  ) $ should have
the same physical content.

But since obviously:
\begin{equation}
     T^{2} ( | E > + | O >  ) \; = \; T^{2} | E > + T^{2} | O > \;
     = \; | E > - | O >
\end{equation}
this is not possible, and hence superposition of states with
different parity under double time-reversal should be banned.

The status of such an argument is not so obvious (for arguments
against it differing by the ones we are going to present we demand
to the $ 11^{th} $ chapter "Superselection Rules" of
\cite{Aharonov-Rohrlich-05}) as we will now show.

Let us start, at this purpose, to review the behavior under
time-reversal of a generic quantum angular momentum $ \vec{J} $:
\begin{equation} \label{eq:angular momentum is odd under time-reversal}
     T \vec{J} T^{-1} \; = \; - \vec{J}
\end{equation}
\begin{equation} \label{eq:action of time-reversal over eigenstates of angular momentum}
    T |j,m > \; = \; i^{2m} |j,-m >
\end{equation}
and hence:
\begin{equation} \label{eq:action of squared time-reversal over eigenstates of angular momentum}
    T^{2} |j,m > \; = \; T  i^{2m} |j,- m > \; = \; i^{-2m} T |j,-m
    > \; = \; i^{-4m}  |j,m > \; = \; (-1)^{2j} |j,m >
\end{equation}
where we have used the fact that:
\begin{equation}
    i^{-4m} \; = \; (-1)^{2j} \; \;  \forall m \in \{ - j , \cdots , j
    \} \, , \,  \forall j \in \frac{\mathbb{N}}{2}
\end{equation}

\smallskip

\begin{remark}
\end{remark}
Let us remark that obviously:
\begin{equation} \label{eq:value of the univalence}
    ( - 1)^{2 j} \; = \; \left\{%
\begin{array}{ll}
    +1 , & \hbox{if $j \in \mathbb{N} $;} \\
    -1, & \hbox{if $ j \in \mathbb{H} $.} \\
\end{array}%
\right.
\end{equation}

\smallskip

So the \emph{double time reversal superselection rule} implies the
\emph{univalence superselection rule}, defined as the
superselection rule having \emph{superselection charge} $ (-1)^{2
j} $ and hence banning the superposition of states with $ j \in
\mathbb{N} $ and states with $ j \in \mathbb{H} $.

\bigskip

Let us now consider the coupling of $ n \in \mathbb{N} : n \geq 2
$ quantum angular momenta $ \vec{J}_{1} , \cdots , \vec{J}_{n} $;
given $ j_{1} , \cdots , j_{n} \in \frac{\mathbb{N}}{2} $ let us
observe that:
\begin{proposition} \label{prop:univalence of coupled angular momenta}
\end{proposition}
\begin{equation}
    ( -1)^{2 j} \; = \; \left\{%
\begin{array}{ll}
    +1 , & \hbox{if $ | \{ j_{i} \in \mathbb{H} \, , \, i \in \{ 1 , \cdots , n \} \} | \in \mathbb{E} $;} \\
    -1, & \hbox{ if  $ | \{ j_{i} \in \mathbb{H} \, , \, i \in \{ 1 , \cdots , n \} \} | \in \mathbb{O} $} \\
\end{array}
\right. \; \; \forall j \in \{ \mathcal{J}_{min} ( j_{1} , \cdots
, j_{n} ) ,  \cdots ,  \mathcal{J}_{max} ( j_{1} , \cdots , j_{n}
)  \}
\end{equation}
\begin{proof}
The thesis follows combining the equation \ref{eq:value of the
univalence} and the proposition \ref{prop:algebra of the integers
union the half-integral}.
\end{proof}

\smallskip
Proposition \ref{prop:univalence of coupled angular momenta} plays
an unexpected rule as to time-reversal properties.

Let us observe, first of all, that:
\begin{equation} \label{eq:time-reversal of angular momenta in the first basis}
    T | j_{1}, m_{1} , \cdots , j_{n}, m_{n} > \; = \; i^{2 \sum_{i=1}^{n}
    m_{i}} | j_{1},-  m_{1} , \cdots , j_{n}, -m_{n} >
\end{equation}
\begin{equation} \label{eq:time-reversal of angular momenta in the second basis}
    T | j_{1}, j_{2}, j_{12} , j_{3} , j_{123} , \cdots , j_{1 \cdots
    n-1} , j_{n} , j, m > \; = \; i^{2m}  | j_{1}, j_{2}, j_{12} , j_{3} , j_{123} , \cdots , j_{1 \cdots
    n-1} , j_{n} , j, -m >
\end{equation}
Applying, from the other side, the time-reversal operator to the
proposition \ref{prop:coupling of n angular momenta} it follows
that:
\begin{multline}
   T | j_{1}, j_{2}, j_{12} , j_{3} , j_{123} , \cdots , j_{1 \cdots
    n-1} , j_{n} , j, m > \; = \\
     T \sum_{m_{1}=-j_{1}}^{j_{1}}
    \sum_{m_{2}=-j_{2}}^{j_{2}} \cdots \sum_{m_{n}=-j_{n}}^{j_{n}}
      C_{j_{1 \cdots n-1}m_{1 \cdots n-1}j_{n}m_{n}}^{jm}   \cdots   C_{j_{12}m_{12}j_{3}m_{3}}^{j_{123}m_{123}}   C_{j_{1}m_{1}j_{2}m_{2}}^{j_{12}m_{12}}
      | j_{1}, m_{1} , \cdots , j_{n} , m_{n} >
\end{multline}
\begin{multline}
    T | j_{1}, j_{2}, j_{12} , j_{3} , j_{123} , \cdots , j_{1 \cdots
    n-1} , j_{n} , j, m > \; = \\
     \sum_{m_{1}=-j_{1}}^{j_{1}} \sum_{m_{2}=-j_{2}}^{j_{2}}
    \sum_{m_{3}=-j_{3}}^{j_{3}} \cdots \sum_{m_{n}=-j_{n}}^{j_{n}}
     T C_{j_{1 \cdots n-1}m_{1 \cdots n-1}j_{n}m_{n}}^{jm}   \cdots   C_{j_{12}m_{12}j_{3}m_{3}}^{j_{123}m_{123}}  C_{j_{1}m_{1}j_{2}m_{2}}^{j_{12}m_{12}}
      | j_{1}, m_{1} , \cdots , j_{n}, m_{n} >
\end{multline}
\begin{multline}
    T | j_{1}, j_{2}, j_{12} , j_{3} , j_{123} , \cdots , j_{1 \cdots
    n-1} , j_{n} , j, m > \; = \\
     \sum_{m_{1}=-j_{1}}^{j_{1}} \sum_{m_{2}=-j_{2}}^{j_{2}} \sum_{m_{3}=-j_{3}}^{j_{3}}
 \cdots \sum_{m_{n}=-j_{n}}^{j_{n}}
    \overline{ C_{j_{1 \cdots n-1}m_{1 \cdots n-1}j_{n}m_{n}}^{jm}   \cdots   C_{j_{12}m_{12}j_{3}m_{3}}^{j_{123}m_{123}}   C_{j_{1}m_{1}j_{2}m_{2}}^{j_{12}m_{12}}}
      T | j_{1}, m_{1} , \cdots , j_{n}, m_{n} >
\end{multline}
where we have used the anti-linearity of T. Hence:
\begin{multline} \label{eq:time-reversal in the Clebsch-Gordan decomposition}
    T | j_{1}, j_{2}, j_{12} , j_{3} , j_{123} , \cdots , j_{1 \cdots
    n-1} , j_{n} , j, m > \; = \\
    \sum_{m_{1}=-j_{1}}^{j_{1}} \sum_{m_{2}=-j_{2}}^{j_{2}}
    \sum_{m_{3}=-j_{3}}^{j_{3}} \cdots  \sum_{m_{n}=-j_{n}}^{j_{n}}
      C_{j_{1 \cdots n-1}m_{1 \cdots n-1}j_{n}m_{n}}^{jm}   \cdots   C_{j_{12}m_{12}j_{3}m_{3}}^{j_{123}m_{123}}  C_{j_{1}m_{1}j_{2}m_{2}}^{j_{12}m_{12}}
     i^{2 \sum_{i=1}^{n} m_{i} }  | j_{1}, - m_{1} , \cdots , j_{n}, - m _{n} >
\end{multline}
where in the last passage we have used the equation
\ref{eq:time-reversal of angular momenta in the first basis} and
the fact that the Clebsch-Gordan coefficients are reals.

Comparing the equation \ref{eq:time-reversal of angular momenta in
the second basis} and the equation \ref{eq:time-reversal in the
Clebsch-Gordan decomposition} it follows that:
\begin{multline}
 | j_{1}, j_{2}, j_{12} , j_{3} , j_{123} , \cdots , j_{1 \cdots
    n-1} , j_{n} , j, -m > \; = \\
\sum_{m_{1}=-j_{1}}^{j_{1}} \sum_{m_{2}=-j_{2}}^{j_{2}}
    \sum_{m_{3}=-j_{3}}^{j_{3}} \cdots \sum_{m_{n}=-j_{n}}^{j_{n}}
C_{j_{1 \cdots n-1}m_{1 \cdots n-1}j_{n}m_{n}}^{jm}   \cdots
C_{j_{12}m_{12}j_{3}m_{3}}^{j_{123}m_{123}}
C_{j_{1}m_{1}j_{2}m_{2}}^{j_{12}m_{12}}
     i^{2(\sum_{i=1}^{n} m_{i} -m )}  | j_{1}, - m_{1} , \cdots , j_{n}, - m _{n}
     > \\
     \forall m \in \{ - j , \cdots , j \} , \forall j \in \{| j_{1 \cdots n-1} - j_{n} | , \cdots , j_{1 \cdots n-1}
      +j_{n} \} \\
 \forall j_{1 \cdots n-1} \in \{| j_{1 \cdots n-2} - j_{n-1} | , \cdots , j_{1 \cdots n-2}
      +j_{n-1} \} , \cdots, \forall j_{123} \in \{| j_{12} - j_{3} | , \cdots , j_{12}
      +j_{3} \} , \forall j_{12} \in \{| j_{1} - j_{2} | , \cdots , j_{1}
      +j_{2} \}
\end{multline}
that using the proposition \ref{prop:basic property of the
Clebsch-Gordan coefficient} reduces to:
\begin{proposition} \label{prop:consequence of the application of time-reversal}
\end{proposition}
\begin{multline}
 | j_{1}, j_{2}, j_{12} , j_{3} , j_{123} , \cdots , j_{1 \cdots
    n-1} , j_{n} , j, -m > \; = \\
\sum_{m_{1}=-j_{1}}^{j_{1}} \sum_{m_{2}=-j_{2}}^{j_{2}}
    \sum_{m_{3}=-j_{3}}^{j_{3}} \cdots \sum_{m_{n}=-j_{n}}^{j_{n}}
C_{j_{1 \cdots n-1}m_{1 \cdots n-1}j_{n}m_{n}}^{jm}   \cdots
C_{j_{12}m_{12}j_{3}m_{3}}^{j_{123}m_{123}}
C_{j_{1}m_{1}j_{2}m_{2}}^{j_{12}m_{12}}
      | j_{1}, - m_{1} , \cdots , j_{n}, - m _{n}
     > \\
     \forall m \in \{ - j , \cdots , j \} , \forall j \in \{| j_{1 \cdots n-1} - j_{n} | , \cdots , j_{1 \cdots n-1}
      +j_{n} \} \\
 \forall j_{1 \cdots n-1} \in \{| j_{1 \cdots n-2} - j_{n-1} | , \cdots , j_{1 \cdots n-2}
      +j_{n-1} \} , \cdots, \forall j_{123} \in \{| j_{12} - j_{3} | , \cdots , j_{12}
      +j_{3} \} , \forall j_{12} \in \{| j_{1} - j_{2} | , \cdots , j_{1}
      +j_{2} \}
\end{multline}
\bigskip

Furthermore:
\begin{equation} \label{eq:squared time-reversal of angular momenta in the first basis}
    T^{2} | j_{1}, m_{1} , \cdots , j_{n}, m_{n} > \; = \; ( -1 )^{2 \sum_{i=1}^{n} j_{i} }  | j_{1},  m_{1} , \cdots , j_{n}, m_{n} >
\end{equation}
\begin{equation} \label{eq:squared time-reversal of angular momenta in the second basis}
    T^{2} | j_{1}, j_{2}, j_{12} , j_{3} , j_{123} , \cdots , j_{1 \cdots
    n-1} , j_{n} , j, m > \; = \; (-1)^{2j}  | j_{1}, j_{2}, j_{12} , j_{3} , j_{123} , \cdots , j_{1 \cdots
    n-1} , j_{n} , j, m >
\end{equation}
Applying, from the other side, the square of the time-reversal
operator to the proposition \ref{prop:coupling of n angular
momenta} it follows that:
\begin{multline}
   T^{2} | j_{1}, j_{2}, j_{12} , j_{3} , j_{123} , \cdots , j_{1 \cdots
    n-1} , j_{n} , j, m > \; = \\
    T^{2} \sum_{m_{1}=-j_{1}}^{j_{1}}
    \sum_{m_{2}=-j_{2}}^{j_{2}}
    \sum_{m_{3}=-j_{3}}^{j_{3}}  \cdots \sum_{m_{n}=-j_{n}}^{j_{n}}
      C_{j_{1 \cdots n-1}m_{1 \cdots n-1}j_{n}m_{n}}^{jm}   \cdots   C_{j_{12}m_{12}j_{3}m_{3}}^{j_{123}m_{123}}   C_{j_{1}m_{1}j_{2}m_{2}}^{j_{12}m_{12}}
      | j_{1}, m_{1} , \cdots , j_{n} m_{n} >
\end{multline}
\begin{multline}
    T^{2} | j_{1}, j_{2}, j_{12} , j_{3} , j_{123} , \cdots , j_{1 \cdots
    n-1} , j_{n} , j, m > \; = \\
     \sum_{m_{1}=-j_{1}}^{j_{1}} \sum_{m_{2}=-j_{2}}^{j_{2}}
    \sum_{m_{3}=-j_{3}}^{j_{3}} \cdots \sum_{m_{n}=-j_{n}}^{j_{n}}
     T^{2} C_{j_{1 \cdots n-1}m_{1 \cdots n-1}j_{n}m_{n}}^{jm}   \cdots   C_{j_{12}m_{12}j_{3}m_{3}}^{j_{123}m_{123}}  C_{j_{1}m_{1}j_{2}m_{2}}^{j_{12}m_{12}}
      | j_{1}, m_{1} , \cdots , j_{n}, m_{n} >
\end{multline}
\begin{multline}
    T^{2} | j_{1}, j_{2}, j_{12} , j_{3} , j_{123} , \cdots , j_{1 \cdots
    n-1} , j_{n} , j, m > \; = \\
    \sum_{m_{1}=-j_{1}}^{j_{1}} \sum_{m_{2}=-j_{2}}^{j_{2}}
    \sum_{m_{3}=-j_{3}}^{j_{3}} \cdots  \sum_{m_{n}=-j_{n}}^{j_{n}}
      C_{j_{1 \cdots n-1}m_{1 \cdots n-1}j_{n}m_{n}}^{jm}   \cdots   C_{j_{12}m_{12}j_{3}m_{3}}^{j_{123}m_{123}}  C_{j_{1}m_{1}j_{2}m_{2}}^{j_{12}m_{12}}
     T^{2} | j_{1},  m_{1} , \cdots , j_{n},  m _{n} >
\end{multline}
where in the last passage we have used the linearity of $ T^{2} $.
Hence:
\begin{multline} \label{eq:squared time-reversal in the Clebsch-Gordan decomposition}
    T^{2} | j_{1}, j_{2}, j_{12} , j_{3} , j_{123} , \cdots , j_{1 \cdots
    n-1} , j_{n} , j, m > \; = \\
    \sum_{m_{1}=-j_{1}}^{j_{1}} \sum_{m_{2}=-j_{2}}^{j_{2}}
    \sum_{m_{3}=-j_{3}}^{j_{3}}  \cdots \sum_{m_{n}=-j_{n}}^{j_{n}}
      C_{j_{1 \cdots n-1}m_{1 \cdots n-1}j_{n}m_{n}}^{jm}   \cdots   C_{j_{12}m_{12}j_{3}m_{3}}^{j_{123}m_{123}}  C_{j_{1}m_{1}j_{2}m_{2}}^{j_{12}m_{12}}
     (-1 )^{2 \sum_{i=1}^{n} j_{i} }  | j_{1},  m_{1} , \cdots , j_{n},  m_{n} >
\end{multline}
where we have used the equation \ref{eq:squared time-reversal of
angular momenta in the first basis}.

Comparing the equation \ref{eq:squared time-reversal of angular
momenta in the second basis} and the equation \ref{eq:squared
time-reversal in the Clebsch-Gordan decomposition} it follows
that:
\begin{proposition} \label{prop:consequence of the application of squared time-reversal}
\end{proposition}
\begin{multline}
    | j_{1}, j_{2}, j_{12} , j_{3} , j_{123} , \cdots , j_{1 \cdots
    n-1} , j_{n} , j, m > \; = \\
     \sum_{m_{1}=-j_{1}}^{j_{1}} \sum_{m_{2}=-j_{2}}^{j_{2}}
    \sum_{m_{3}=-j_{3}}^{j_{3}} \cdots \sum_{m_{n}=-j_{n}}^{j_{n}}
      C_{j_{1 \cdots n-1}m_{1 \cdots n-1}j_{n}m_{n}}^{jm}   \cdots   C_{j_{12}m_{12}j_{3}m_{3}}^{j_{123}m_{123}}  C_{j_{1}m_{1}j_{2}m_{2}}^{j_{12}m_{12}}
     (-1 )^{2 (\sum_{i=1}^{n} j_{i} - j) }  | j_{1},  m_{1} , \cdots , j_{n},  m_{n}
     > \\
     \forall m \in \{ - j , \cdots , j \} , \forall j \in \{| j_{1 \cdots n-1} - j_{n} | , \cdots , j_{1 \cdots n-1}
      +j_{n} \} \\
 \forall j_{1 \cdots n-1} \in \{| j_{1 \cdots n-2} - j_{n-1} | , \cdots , j_{1 \cdots n-2}
      +j_{n-1} \} , \cdots, \forall j_{123} \in \{| j_{12} - j_{3} | , \cdots , j_{12}
      +j_{3} \} , \forall j_{12} \in \{| j_{1} - j_{2} | , \cdots , j_{1}
      +j_{2} \}
\end{multline}

\bigskip

Let us now observe that while no compatibility problem exists
between the proposition \ref{prop:coupling of n angular momenta}
and the proposition \ref{prop:consequence of the application of
time-reversal}, the compatibility between the proposition
\ref{prop:coupling of n angular momenta} and the proposition
\ref{prop:consequence of the application of squared time-reversal}
requires that:
\begin{equation} \label{eq:compatibility condition}
     (-1 )^{2 (\sum_{i=1}^{n} j_{i} - j) } \; = \; 1 \; \;
      \forall j \in \{ \mathcal{J}_{min}( j_{1} , \cdots , j_{n} ) , \cdots , \mathcal{J}_{max}( j_{1} , \cdots , j_{n} )  \} , \forall j_{1} ,
     \cdots , j_{n} \in \frac{\mathbb{N}}{2}
\end{equation}

But equation \ref{eq:compatibility condition} is a trivial
consequence of the proposition \ref{prop:univalence of coupled
angular momenta}.

\bigskip

\begin{remark} \label{rem:irrilevance of orbital angular momentum}
\end{remark}
Let us now observe that the angular momentum $ \vec{J} $ of a
particle is obtained coupling its \emph{orbital angular momentum}
$ \vec{L} $ and its \emph{spin} $ \vec{S} $.

The particle is defined to be a \emph{boson} if its univalence is
equal to $ +1 $ while it is defined to be a \emph{fermion} if its
univalence is equal to $ -1 $.

Since the \emph{univalence} of the \emph{orbital angular momentum}
is always equal to +1, the  proposition \ref{prop:univalence of
coupled angular momenta} implies that the univalence of the
particle is equal to the univalence of its spin.

\bigskip

\begin{remark} \label{rem:classification of the particles}
\end{remark}
The remark \ref{rem:irrilevance of orbital angular momentum}
allows to give the following characterization of bosons and
fermions:
\begin{enumerate}
    \item a \emph{fermion} is a particle containing an odd number of
    \emph{fermions}
    \item a \emph{boson} is a particle containing an even number of
   \emph{ fermions} \footnote{Let us recall that zero is an even number.}
\end{enumerate}
that, restricting the analysis to particles composed by a finite
number of "basic particles" of known univalence \footnote{Let us
remark that we have not assumed that these "basic particles" are
elementary; we could choose as \emph{"basic particles}", for
instance, the atom of $ He_{4} $ (known to be a \emph{boson}), and
the atom of $He_{3}$ (known to be a \emph{fermion}) arriving to
the same conclusion, as to the univalence's attribution, that we
would have reached if we had chosen as "basic particles" the
proton, the neutron and the electron (all known to be fermions)
and we had inferred that the atom of $He_{3}$ is a fermion by the
fact that it is composed by 1 neutron, 2 protons and 2 electrons
and we had inferred that the atom of $He_{4}$ is a boson by the
fact that it is composed by 2 neutrons, 2 protons and 2 neutrons,
or if we had chosen as "basic particles" the electron, the quark
up and the quark down (all known to be fermions) and, considering
that the proton is composed by 2 quarks up and 1 quark down while
the neutron is composed by 1 quark up and 2 quarks down,  we had
inferred that the atom of $He_{3}$ is a fermion by the fact that
it is composed by 2 electrons, 5 quarks up and 4 quarks down and
we had inferred that the atom of $He_{4}$ is a boson by the fact
that it is composed by 2 electrons, 6 quarks up and 6 quarks
down.} and representing mathematically the predicate \emph{"being
a subparticle"} with the set-theoretic membership relation , may
be implemented through the following Mathematica expression
\verb"fermionQ[x]" (where \verb"x" has to be a list whose atomic
elements are the univalences of the "basic particles"):
\begin{verbatim}
fermionQ[x_] := If[AtomQ[x], x == -1, OddQ[Length[Select[x,
fermionQ]]]]
\end{verbatim}

\smallskip

\begin{remark}
\end{remark}
Let us now observe that since the univalence of a compound
particle is, as stated by the remark \ref{rem:classification of
the particles}, already determined by the Coupling Theory, the
\emph{double time-reversal superselection rule} is redundant.

\newpage
\section{Symmetries of the Clebsch-Gordan coefficients deducible from time-reversal
considerations} \label{sec:Symmetries of the Clebsch-Gordan
coefficients deducible from time-reversal considerations}

In the previous section we have actually used only the
anti-linearity of T while we have never used its anti-unitarity.

Resorting to this property allows, indeed, to derive some symmetry
properties of the Clebsch-Gordan coefficients.

Given a state $ | \psi > $ let us introduce the following useful
and suggestive (one can look at the overleft arrow as the inverted
arrow of time) notation:
\begin{equation}
  \overleftarrow{  | \psi >    } \; := \; T  | \psi >
\end{equation}

Considering again the setting of the previous section, let us then
rewrite the equation \ref{eq:time-reversal of angular momenta in
the first basis} as:
\begin{equation} \label{eq:time-reversal of angular momenta in the first basis in new style}
     \overleftarrow{| j_{1}, m_{1} , \cdots , j_{n}, m_{n} > }\; = \; i^{2 \sum_{i=1}^{n}
    m_{i}} | j_{1},-  m_{1} , \cdots , j_{n}, -m_{n} >
\end{equation}
and let us rewrite the equation \ref{eq:time-reversal of angular
momenta in the second basis} as:
\begin{equation} \label{eq:time-reversal of angular momenta in the second basis  in new style}
    \overleftarrow{ | j_{1}, j_{2}, j_{12} , j_{3} , j_{123} , \cdots , j_{1 \cdots
    n-1} , j_{n} , j, m >} \; = \; i^{2m}  | j_{1}, j_{2}, j_{12} , j_{3} , j_{123} , \cdots , j_{1 \cdots
    n-1} , j_{n} , j, -m >
\end{equation}
Hence:
\begin{multline}
  \overleftarrow{ < j_{1}, m_{1} , \cdots , j_{n}, m_{n} } \overleftarrow{ | j_{1}, j_{2}, j_{12} , j_{3} , j_{123} , \cdots , j_{1 \cdots
    n-1} , j_{n} , j, m >} \; = \\
    = i^{2 ( m -  \sum_{i=1}^{n} m_{i})}  < j_{1}, - m_{1} , \cdots , j_{n}, -m_{n} | j_{1}, j_{2},
j_{12} , j_{3} , j_{123} , \cdots , j_{1 \cdots
    n-1} , j_{n} , j, -m >
\end{multline}
By the anti-unitarity of T it follows that:
\begin{multline}
     \overleftarrow{ < j_{1}, m_{1} , \cdots , j_{n}, m_{n} } \overleftarrow{ | j_{1}, j_{2}, j_{12} , j_{3} , j_{123} , \cdots , j_{1 \cdots
    n-1} , j_{n} , j, m >} \; = \\
     \overline{ < j_{1}, m_{1} , \cdots , j_{n}, m_{n} | j_{1}, j_{2}, j_{12} , j_{3} , j_{123} , \cdots , j_{1
\cdots
    n-1} , j_{n} , j, m >}
\end{multline}
and hence:
\begin{multline} \label{eq:symmetry property in Dirac notation}
    \overline{ < j_{1}, m_{1} , \cdots , j_{n},
    m_{n} | j_{1}, j_{2}, j_{12} , j_{3} , j_{123} , \cdots , j_{1 \cdots
    n-1} , j_{n} , j, m >}  \; = \\
     i^{2 ( m -  \sum_{i=1}^{n}
    m_{i})}  < j_{1}, - m_{1} , \cdots , j_{n}, -m_{n} | j_{1}, j_{2}, j_{12} , j_{3} , j_{123} , \cdots , j_{1 \cdots
    n-1} , j_{n} , j, -m >
\end{multline}
Since by the proposition \ref{prop:coupling of n angular momenta}:
\begin{multline}
    < j_{1}, m_{1} , \cdots , j_{n},
    m_{n} | j_{1}, j_{2}, j_{12} , j_{3} , j_{123} , \cdots , j_{1 \cdots
    n-1} , j_{n} , j, m > \; = \\
      C_{j_{1 \cdots n-1}m_{1 \cdots n-1}j_{n}m_{n}}^{jm}   \cdots   C_{j_{12}m_{12}j_{3}m_{3}}^{j_{123}m_{123}}   C_{j_{1}m_{1}j_{2}m_{2}}^{j_{12}m_{12}}
\end{multline}
the equation \ref{eq:symmetry property in Dirac notation} may be
rewritten as:
\begin{multline}
     \overline{C_{j_{1 \cdots n-1}m_{1 \cdots n-1}j_{n}m_{n}}^{jm}   \cdots   C_{j_{12}m_{12}j_{3}m_{3}}^{j_{123}m_{123}}
     C_{j_{1}m_{1}j_{2}m_{2}}^{j_{12}m_{12}}} \; = \\
i^{2 ( m -  \sum_{i=1}^{n}
    m_{i})} C_{j_{1 \cdots n-1}, -m_{1 \cdots n-1},j_{n}, -m_{n}}^{j, -m}   \cdots   C_{j_{12},-m_{12},j_{3},-m_{3}}^{j_{123},-m_{123}}   C_{j_{1}, -m_{1},j_{2},- m_{2}}^{j_{12}, -m_{12}}
\end{multline}
(where we have used commas to separate the indices in order to
avoid the notational ambiguity resulting by the minus signs) that
using the reality of the Clebsch-Gordan coefficients gives:
\begin{multline}
     C_{j_{1 \cdots n-1}m_{1 \cdots n-1}j_{n}m_{n}}^{jm}   \cdots   C_{j_{12}m_{12}j_{3}m_{3}}^{j_{123}m_{123}}
     C_{j_{1}m_{1}j_{2}m_{2}}^{j_{12}m_{12}} \; = \\
i^{2 ( m -  \sum_{i=1}^{n}
    m_{i})} C_{j_{1 \cdots n-1}, -m_{1 \cdots n-1},j_{n}, -m_{n}}^{j, -m}   \cdots   C_{j_{12},-m_{12},j_{3},-m_{3}}^{j_{123},-m_{123}}   C_{j_{1}, -m_{1},j_{2},- m_{2}}^{j_{12},
    -m_{12}} \\
    \forall m_{1} \in \{ - j_{1} , \cdots , j_{1} \} , \cdots , \forall m_{n} \in \{ - j_{n} , \cdots , j_{n}
    \} ,
    \forall m \in \{ - j , \cdots , j \} , \\
     \forall j \in \{| j_{1 \cdots n-1} - j_{n} | , \cdots , j_{1 \cdots n-1}
      +j_{n} \},
 \forall j_{1 \cdots n-1} \in \{| j_{1 \cdots n-2} - j_{n-1} | , \cdots , j_{1 \cdots n-2}
      +j_{n-1} \} , \cdots , \\
       \forall j_{123} \in \{| j_{12} - j_{3} | , \cdots , j_{12}
      +j_{3} \} , \forall j_{12} \in \{| j_{1} - j_{2} | , \cdots , j_{1}
      +j_{2} \} , \forall j_{1} , \cdots , j_{n} \in \frac{\mathbb{N}}{2}
\end{multline}

Using the proposition \ref{prop:basic property of the
Clebsch-Gordan coefficient} it follows that:
\begin{proposition}  \label{prop:first symmetry property}
\end{proposition}
\begin{multline}
     C_{j_{1 \cdots n-1}m_{1 \cdots n-1}j_{n}m_{n}}^{jm}   \cdots   C_{j_{12}m_{12}j_{3}m_{3}}^{j_{123}m_{123}}
     C_{j_{1}m_{1}j_{2}m_{2}}^{j_{12}m_{12}} \; = \\
 C_{j_{1 \cdots n-1}, -m_{1 \cdots n-1},j_{n}, -m_{n}}^{j, -m}
\cdots C_{j_{12},-m_{12},j_{3},-m_{3}}^{j_{123},-m_{123}}
C_{j_{1}, -m_{1},j_{2},- m_{2}}^{j_{12},
    -m_{12}} \\
    \forall m_{1} \in \{ - j_{1} , \cdots , j_{1} \} , \cdots , \forall m_{n} \in \{ - j_{n} , \cdots , j_{n}
    \} ,
    \forall m \in \{ - j , \cdots , j \} , \\
     \forall j \in \{| j_{1 \cdots n-1} - j_{n} | , \cdots , j_{1 \cdots n-1}
      +j_{n} \},
 \forall j_{1 \cdots n-1} \in \{| j_{1 \cdots n-2} - j_{n-1} | , \cdots , j_{1 \cdots n-2}
      +j_{n-1} \} , \cdots, \\
       \forall j_{123} \in \{| j_{12} - j_{3} | , \cdots , j_{12}
      +j_{3} \} , \forall j_{12} \in \{| j_{1} - j_{2} | , \cdots , j_{1}
      +j_{2} \} , \forall j_{1} , \cdots , j_{n} \in \frac{\mathbb{N}}{2}
\end{multline}

\smallskip

Let us now recall the following:

\begin{proposition} \label{prop:basic property of states odd under squared time-reversal}
\end{proposition}
\emph{Basic property of states odd under double time reversal:}
\begin{equation}
   \overleftarrow{\overleftarrow{ | \psi >}} \, = \, - | \psi > \;
   \Rightarrow \;\overleftarrow{ < \psi |} \psi >  \, = \, 0
\end{equation}
\begin{proof}
Since T is antiunitary:
\begin{equation} \label{eq:antiunitary of time-reversal}
  \overleftarrow{< \alpha |}\overleftarrow{ \beta >} \; = \; \overline{< \alpha
  | \beta >} \; = \; < \beta | \alpha > \; \; \forall | \alpha > ,
  | \beta >
\end{equation}
Choosing in particular $  | \alpha > \, := \  | \psi >  $ and $ |
\beta > \, := \,\overleftarrow{ | \psi >} $ the  equation
\ref{eq:antiunitary of time-reversal} gives:
\begin{equation}
   \overleftarrow{ < \psi |} \psi > \; = \;  \overleftarrow{ < \psi
   }\overleftarrow{\overleftarrow{ | \psi >}}
\end{equation}
that using the hypothesis that $ | \psi > $ is odd under double
time-reversal becomes:
\begin{equation}
  \overleftarrow{ < \psi |} \psi > \; = \; - \overleftarrow{ < \psi |} \psi >
\end{equation}
from which the thesis follows.
\end{proof}

\smallskip

Applying the proposition \ref{prop:basic property of states odd
under squared time-reversal} to the state $| j_{1}, j_{2}, j_{12}
, j_{3} , j_{123} , \cdots , j_{1 \cdots
    n-1} , j_{n} , j, m > $ with $ j \in \mathbb{H} $ (where,
    following the notation introduced in the section \ref{sec:Some useful algebraic
    properties}, $ \mathbb{H} $ is the set of the \emph{
    half-odd numbers}), we obtain:
\begin{equation}
 < j_{1}, j_{2}, j_{12}
, j_{3} , j_{123} , \cdots , j_{1 \cdots
    n-1} , j_{n} , j, m  \overleftarrow{| j_{1}, j_{2}, j_{12}
, j_{3} , j_{123} , \cdots , j_{1 \cdots
    n-1} , j_{n} , j, m >} \; = \; 0
\end{equation}
that, using the proposition \ref{prop:coupling of n angular
momenta} and the equation \ref{eq:time-reversal in the
Clebsch-Gordan decomposition}, gives:

\begin{multline} \label{eq:using the basic property to recoupling1}
     \sum_{m_{1}=-j_{1}}^{j_{1}}
    \sum_{m_{2}=-j_{2}}^{j_{2}} \cdots \sum_{m_{n}=-j_{n}}^{j_{n}}
     \overline{ C_{j_{1 \cdots n-1}m_{1 \cdots n-1}j_{n}m_{n}}^{jm}   \cdots   C_{j_{12}m_{12}j_{3}m_{m}}^{j_{123}m_{3}}   C_{j_{1}m_{1}j_{2}m_{2}}^{j_{12}m_{12}}}
      < j_{1}, m_{1} , \cdots , j_{n}, m _{n} | \\
      \sum_{m'_{1}=-j_{1}}^{j_{1}} \sum_{m'_{2}=-j_{2}}^{j_{2}}
    \sum_{m'_{3}=-j_{3}}^{j_{3}} \cdots  \sum_{m'_{n}=-j_{n}}^{j_{n}}
      C_{j_{1 \cdots n-1}m'_{1 \cdots n-1}j_{n}m'_{n}}^{jm'}   \cdots   C_{j_{12}m'_{12}j_{3}m'_{3}}^{j_{123}m'_{123}}  C_{j_{1}m'_{1}j_{2}m'_{2}}^{j_{12}m'_{12}}
     i^{2 \sum_{i=1}^{n} m'_{i} }  | j_{1}, - m'_{1} , \cdots , j_{n}, - m' _{n} > \; = \; 0
\end{multline}
Since:
\begin{equation}
    < j_{1}, m_{1} , \cdots , j_{n}, m _{n} | j_{1}, - m'_{1} , \cdots , j_{n}, - m' _{n}
    > \; = \; \prod_{i=1}^{n} \delta_{m_{i},- m'_{i}}
\end{equation}
(and using the reality of the Clebsch-Gordan coefficients) the
only non-zero addends in the left hand side of the equation
\ref{eq:using the basic property to recoupling1} are those with $
m'_{1}= - m_{1} , \cdots , m'_{n}= - m_{n}$  for which one
obtains:
\begin{proposition}  \label{prop:second symmetry property}
\end{proposition}
\begin{multline}
     \sum_{m_{1}=-j_{1}}^{j_{1}}
    \sum_{m_{2}=-j_{2}}^{j_{2}} \cdots \sum_{m_{n}=-j_{n}}^{j_{n}}
     \\
    i^{- 2 (\sum_{i=1}^{n} m_{i}) }   C_{j_{1 \cdots n-1}m_{1 \cdots n-1}j_{n}m_{n}}^{jm}   \cdots   C_{j_{12}m_{12}j_{3}m_{3}}^{j_{123}m_{123}}   C_{j_{1}m_{1}j_{2}m_{2}}^{j_{12}m_{12}}
C_{j_{1 \cdots n-1},-m_{1 \cdots n-1}j_{n},-m_{n}}^{j,-m}   \cdots
C_{j_{12},-m_{12},j_{3},-m_{3}}^{j_{123},-m_{123}}
C_{j_{1},-m_{1},j_{2},-m_{2}}^{j_{12},-m_{12}}
 \; = \; 0 \\
 \forall m_{1} \in \{ - j_{1} , \cdots , j_{1} \} , \cdots , \forall m_{n} \in \{ - j_{n} , \cdots , j_{n}
    \} ,
    \forall m \in \{ - j , \cdots , j \} , \\
     \forall j \in \{| j_{1 \cdots n-1} - j_{n} | , \cdots , j_{1 \cdots n-1}
      +j_{n} \} : j \in \mathbb{H} ,
 \forall j_{1 \cdots n-1} \in \{| j_{1 \cdots n-2} - j_{n-1} | , \cdots , j_{1 \cdots n-2}
      +j_{n-1} \} , \cdots , \\
       \forall j_{123} \in \{| j_{12} - j_{3} | , \cdots , j_{12}
      +j_{3} \} , \forall j_{12} \in \{| j_{1} - j_{2} | , \cdots , j_{1}
      +j_{2} \} , \forall j_{1} , \cdots , j_{n} \in \frac{\mathbb{N}}{2}
\end{multline}

Let us now compare the proposition \ref{prop:first symmetry
property} and the proposition \ref{prop:second symmetry property}
with the symmetry properties of the Clebsch-Gordan coefficients
(discovered by Tullio Regge in 1954 \cite{Regge-58}).

Let us introduce at this purpose the \emph{Wigner 3j symbol}:
\begin{equation}
    \left(%
\begin{array}{ccc}
  j_{1} & j_{2} & j \\
  m_{1} & m_{2} & m \\
\end{array}%
\right) \; := \;  \frac{(-1)^{j_{1}-j_{2}+m}}{\sqrt{2j+1} }
C_{j_{1},m_{1},j_{2},m_{2}}^{j,-m}
\end{equation}
in terms of which the \emph{Regge R-symbol}:
\begin{equation}
  \| \begin{array}{ccc}
    R_{11} & R_{12} & R_{13} \\
    R_{21} & R_{22} & R_{23} \\
    R_{31} & R_{32} & R_{33} \\
  \end{array}    \| \; := \; \left(%
\begin{array}{ccc}
  a & b & c \\
  \alpha & \beta & \gamma \\
\end{array}%
\right)
\end{equation}
is defined by the relations:
\begin{equation}
    \begin{array}{ccc}
      R_{11} := -a + b+ c & R_{12} := a - b +c & R_{13} := a + b - c \\
      R_{21} := a + \alpha & R_{22} := b + \beta & R_{33} := c + \gamma \\
      R_{31} := a - \alpha & R_{32} := b - \beta  & R_{33} := c - \gamma \\
    \end{array}
\end{equation}

Then the symmetries of the Clebsch-Gordan coefficients are encoded
in the following:

\begin{proposition} \label{prop:symmetry property of the R-symbol}
\end{proposition}
\begin{enumerate}
    \item permutation of the rows:
\begin{equation}
\| \begin{array}{ccc}
    R_{11} & R_{12} & R_{13} \\
    R_{21} & R_{22} & R_{23} \\
    R_{31} & R_{32} & R_{33} \\
  \end{array}    \| \; = \; \epsilon_{ijk}
  \| \begin{array}{ccc}
    R_{i1} & R_{i2} & R_{i3} \\
    R_{j1} & R_{j2} & R_{j3} \\
    R_{k1} & R_{k2} & R_{k3} \\
  \end{array}    \| \; \; \forall i,j,k \in \{ 1,2,3 \}
\end{equation}
    \item permutation of the columns:
\begin{equation}
    \| \begin{array}{ccc}
    R_{11} & R_{12} & R_{13} \\
    R_{21} & R_{22} & R_{23} \\
    R_{31} & R_{32} & R_{33} \\
  \end{array}    \| \; = \; \epsilon_{ijk} \| \begin{array}{ccc}
    R_{1i} & R_{1j} & R_{1k} \\
    R_{2i} & R_{2j} & R_{2k} \\
    R_{3i} & R_{3j} & R_{3k} \\
  \end{array}    \| \; \; \forall i,j,k \in \{ 1,2,3 \}
\end{equation}
    \item transposition:
\begin{equation}
    \| \begin{array}{ccc}
    R_{11} & R_{12} & R_{13} \\
    R_{21} & R_{22} & R_{23} \\
    R_{31} & R_{32} & R_{33} \\
  \end{array}    \| \; = \; \| \begin{array}{ccc}
    R_{11} & R_{21} & R_{31} \\
    R_{12} & R_{22} & R_{32} \\
    R_{13} & R_{23} & R_{33} \\
  \end{array}  \|
\end{equation}
\end{enumerate}

Actually the proposition \ref{prop:first symmetry property} may be
derived simply by the proposition \ref{prop:symmetry property of
the R-symbol} applying to each of the Clebsch-Gordan coefficients
appearing in it its transformation property under change of sign
of all the angular momentum projection's (corresponding to the
permutation of the second and the third rows of the R-symbol).

\bigskip

We strongly suspect  that also the proposition \ref{prop:second
symmetry property} can be derived from the proposition
\ref{prop:symmetry property of the R-symbol} though at the present
time we don't know how.

\newpage
\section{Kramers degenerations, hidden symmetries and Coupling
Theory} \label{sec:Kramers degenerations, Hidden Symmetries and
Coupling Theory}

Proposition \ref{prop:basic property of states odd under squared
time-reversal} was used by Wigner to show that the degeneracy of
the energy levels of an odd number of electrons subjected to an
arbitrary electric field could be derived from first principles as
a particular case of the following:
\begin{theorem} \label{th:Wigner's Theorem about Kramers degeneracy}
\end{theorem}
\emph{Wigner's Theorem about Kramers degeneracy:}

\begin{hypothesis}
\end{hypothesis}
\begin{center}
 H hamiltonian of a quantum system such that $ [ H , T ] = 0 $
\end{center}
\begin{equation}
    H | \psi > \; = \; E  | \psi >
\end{equation}
\begin{equation}
    \overleftarrow{\overleftarrow{ | \psi >}} \, = \, - | \psi >
\end{equation}
\begin{thesis}
\end{thesis}
\begin{equation}
    degeneration(E) \; \geq \; 2
\end{equation}
\begin{proof}
\begin{equation}
    H \overleftarrow{| \psi >} \; = \; \overleftarrow{H  | \psi
    > } \; = \;   \overleftarrow{E | \psi > }\; = \; \bar{E}  \overleftarrow{| \psi > }
\end{equation}
Since H is self-adjoint it follows that $ E \in \mathbb{R} $ and
hence:
\begin{equation}
    H \overleftarrow{| \psi >} \; = \; E   \overleftarrow{| \psi >}
\end{equation}
Since by the proposition \ref{prop:basic property of states odd
under squared time-reversal}:
\begin{equation}
    < \psi \overleftarrow{ | \psi > } \; = \; 0
\end{equation}
the thesis follows.
\end{proof}

\bigskip

Let us now suppose to have  a system of $ n \in \mathbb{N} : n
\geq 2 $ quantum angular momenta $ \vec{J}_{1} , \cdots ,
\vec{J}_{n}$ with an hamiltonian H having time-reversal and
rotational symmetry, i.e. such that:
\begin{equation}
  [ H , T] \; = \;  [ H , \vec{J}^{2} ] \; = \;  [ H, J_{3} ] =  \; 0
\end{equation}
Then:
\begin{equation}
    H | \alpha ,  j_{1}, j_{2}, j_{12} , j_{3} , j_{123} , \cdots , j_{1 \cdots
    n-1} , j_{n} , j, m >  \; = \; E_{\alpha,j}   | \alpha ,  j_{1}, j_{2}, j_{12} , j_{3} , j_{123} , \cdots , j_{1 \cdots
    n-1} , j_{n} , j, m >
\end{equation}
\begin{equation}
    \vec{J}^{2}  | \alpha ,  j_{1}, j_{2}, j_{12} , j_{3} , j_{123} , \cdots , j_{1 \cdots
    n-1} , j_{n} , j, m >  \; = \; j(j+1)  | \alpha ,  j_{1}, j_{2}, j_{12} , j_{3} , j_{123} , \cdots , j_{1 \cdots
    n-1} , j_{n} , j, m >
\end{equation}
\begin{equation}
    J_{3}   | \alpha ,  j_{1}, j_{2}, j_{12} , j_{3} , j_{123} , \cdots , j_{1 \cdots
    n-1} , j_{n} , j, m > \; = \; m   | \alpha ,  j_{1}, j_{2}, j_{12} , j_{3} , j_{123} , \cdots , j_{1 \cdots
    n-1} , j_{n} , j, m >
\end{equation}
\begin{equation}
     \overleftarrow{ | \alpha ,  j_{1}, j_{2}, j_{12} , j_{3} , j_{123} , \cdots , j_{1 \cdots
    n-1} , j_{n} , j, m >} \; = \; i^{2m}  |\overleftarrow{ \alpha },  j_{1}, j_{2}, j_{12} , j_{3} , j_{123} , \cdots , j_{1 \cdots
    n-1} , j_{n} , j,-  m >
\end{equation}
\begin{equation}
    \overleftarrow{\overleftarrow{ | \alpha ,  j_{1}, j_{2}, j_{12} , j_{3} , j_{123} , \cdots , j_{1 \cdots
    n-1} , j_{n} , j, m > }} \; = \; ( - 1)^{2 j} | \overleftarrow{\overleftarrow{\alpha}} ,  j_{1}, j_{2}, j_{12} , j_{3} , j_{123} , \cdots , j_{1 \cdots
    n-1} , j_{n} , j, m >
\end{equation}
where we have denoted with the label $ \alpha $ a suitable set of
quantum numbers encoding every suppletive hidden symmetry.

Given $ j \in \mathbb{H} $ let us observe that theorem
\ref{th:Wigner's Theorem about Kramers degeneracy} can be applied
if and only if $ \overleftarrow{\overleftarrow{ \alpha }} \, = \,
\alpha $.

\smallskip

\begin{remark}
\end{remark}
Let us remark that the presence of hidden symmetries can lead to
surprises  both in implementing a \emph{coupling scheme} and in
Recoupling Theory where the transition between different
\emph{coupling schemes} is implemented, as we will show through
some examples.

\smallskip

\begin{example} \label{ex:Z bosons in a Keplerian field}
\end{example}
Let us consider  $ Z \in \mathbb{N} : n \geq 2 $ bosons of spin
zero and unary mass subjected to a force's field with Keplerian
energy potential $ V(r) := - \frac{1}{r} $ and hence having
hamiltonian:
\begin{equation}
    H \; :=  \; \sum_{I=1}^{Z} H_{I}
\end{equation}
\begin{equation}
    H_{I} \; := \; \frac{ \vec{p}_{I}^{2}}{ 2  } - \frac{1}{|\vec{x}_{I}|} \; \; I \in \{ 1 , \cdots , Z \}
\end{equation}

As it is well-known the Keplerian energy potential has a symmetry
group  bigger than the rotational symmetry  group SO(3) of a
generic central force's field; specifically such a symmetry group
(easily obtained considering that, beside the angular momentum $
\vec{L} $, there is a suppletive conserved quantity: the
Laplace-Runge-Lenz vector) is:
\begin{itemize}
    \item SO(4) for bound states
    \item SO(3,1) for free states
\end{itemize}
and is responsible of the closeness of the classical  orbits (cfr.
for instance the $ 14^{th} $ chapter "Dynamical Symmetries" of
\cite{Greiner-Muller-94}).

Let us introduce the Laplace-Runge-Lenz operators:
\begin{equation}
    \vec{M}_{I} \; := \; \frac{1}{2 } ( \vec{p}_{I} \wedge \vec{L}_{I}
    - \vec{L}_{I} \wedge \vec{p}_{I} ) -  \frac{\vec{x}_{I}}{ | \vec{x}_{I} |
    } \; \;  I \in \{ 1 , \cdots , Z \}
\end{equation}
where of course $ \vec{L}_{I} := \vec{x}_{I} \wedge \vec{p}_{I} $
is the orbital angular momentum of the $ I^{th} $ electron.

Given an eigenvalue $ E_{I}  $ corresponding to a bound state of $
H_{I} $ let us introduce the rescaled  Laplace-Runge-Lenz
operators:
\begin{equation} \label{eq:rescaled Laplace-Runge-Lenz operators}
  \vec{M}_{I}' \; := \; \sqrt{ - \frac{1}{2 E_{I}}}  \vec{M}_{I}
   \; \;  I \in \{ 1 , \cdots , Z \}
\end{equation}
The Lie algebra generated by $ \{ \vec{L}_{I} ,  \vec{M}_{I}'
\}_{I=1}^{Z} $ is completely determined by the following
relations:
\begin{equation} \label{eq:first relation defining the Lie algebra}
    [ L_{Ii} , L_{Jj} ] \; = \; i \delta_{IJ}
    \epsilon_{ijk} L_{Ik} \; \; \forall I,J \in \{ 1 , \cdots , Z
    \} , \forall i,j \in \{1,2,3 \}
\end{equation}
\begin{equation} \label{eq:second relation defining the Lie algebra}
    [ M'_{Ii} , M'_{Jj} ] \; = \; i \delta_{IJ}
    \epsilon_{ijk} L_{Ik} \; \; \forall I,J \in \{ 1 , \cdots , Z
    \} , \forall i,j \in \{1,2,3 \}
\end{equation}
\begin{equation} \label{eq:third relation defining the Lie algebra}
    [ M'_{Ii} , L_{Jj} ] \; = \; i \delta_{IJ}
    \epsilon_{ijk} M'_{Ik} \; \; \forall I,J \in \{ 1 , \cdots , Z
    \} , \forall i,j \in \{1,2,3 \}
\end{equation}
that together with:
\begin{equation}
    [ H, L_{Ii} ] \; = \;  [ H, M'_{Ii} ] \; = \; 0 \; \; \forall I,J \in \{ 1 , \cdots , Z
    \} , \forall i \in \{1,2,3 \}
\end{equation}
describe the $ \otimes_{I=1}^{Z} SO(4) $ symmetry of H.

Introduced the operators:
\begin{equation} \label{eq:first angular momentum operator}
    \vec{J}_{(1)I} \; := \; \frac{1}{2} ( \vec{L}_{I} +
    \vec{M}'_{I}) \; \; I \in \{ 1 , \cdots , Z
    \}
\end{equation}
\begin{equation} \label{eq:second angular momentum operator}
    \vec{J}_{(2)I} \; := \; \frac{1}{2} ( \vec{L}_{I} -
    \vec{M}'_{I}) \; \; I \in \{ 1 , \cdots , Z
    \}
\end{equation}
the equation \ref{eq:first relation defining the Lie algebra}, the
equation \ref{eq:second relation defining the Lie algebra} and the
equation \ref{eq:third relation defining the Lie algebra} imply
that:
\begin{equation} \label{eq:compact form of the Lie algebra}
    [ J_{(\alpha)Ii} ,  J_{(\beta)Jj} ] \; = \; i \delta_{\alpha
    \beta} \delta_{IJ}
    \epsilon_{ijk} J_{(\alpha)Ik} \; \; \forall \alpha , \beta \in
    \{ 1 , 2 \}, \forall I,J \in \{ 1 , \cdots , Z
    \} , \forall i,j \in \{1,2,3 \}
\end{equation}
\begin{equation} \label{eq:symmetry of the hamiltonian}
    [ H , J_{(\alpha)Ii} ] \; = \; 0 \; \; \forall \alpha  \in
    \{ 1 , 2 \}, \forall I \in \{ 1 , \cdots , Z
    \} , \forall i\in \{1,2,3 \}
\end{equation}
The rank \footnote{defined as the maximal number of commuting
generators.} of $ \otimes_{I=1}^{Z} SO(4) $ is clearly 2 Z.

Let us now recall the following basic theorem (see for instance
the section 3.6 "Theorem of Racah" and the chapter 16 "Proof of
Racah's Theorem" of \cite{Greiner-Muller-94}) proved by Giulio
Racah in 1950:

\begin{theorem} \label{th:Racah's theorem}
\end{theorem}
\emph{Racah's Theorem:}
\begin{center}
 The number of linearly independent Casimir operators of a
 semisimple Lie group is equal to its rank
\end{center}

According to the theorem \ref{th:Racah's theorem}   it follows
that $ \otimes_{I=1}^{Z} SO(4) $ has 2 Z linearly independent
Casimir operators that may be chosen to be  $ \{
\vec{J}_{(1)I}^{2} , \vec{J}_{(2)I}^{2} \; I \in \{ 1 , \cdots , Z
\} \} $ whose eigenvalues may be used to characterize the
multiplets \footnote{defined as the irreducible invariant
subspaces.}.

Since the equation \ref{eq:compact form of the Lie algebra} shows
that such operators are 2 Z uncoupled quantum angular momenta it
follows that:
\begin{multline}
    \vec{J}_{(1)I}^{2} | j_{(1),1}, m_{(1),1}, \cdots , j_{(1),Z},
    m_{(1),Z} , j_{(2),1}, m_{(2),1}, \cdots , j_{(2),Z},
    m_{(2),Z} > \; = \\
     j_{(1)I} ( j_{(1)I} + 1 ) | j_{(1),1}, m_{(1),1}, \cdots , j_{(1),Z},
    m_{(1),Z} , j_{(2),1}, m_{(2),1}, \cdots , j_{(2),Z},
    m_{(2),Z} > \\
     \forall  m_{(i),I} \in \{ - j_{(i)I} , \cdots , j_{(i)I} \} \, , \, \forall j_{(i)I} \in \frac{\mathbb{N}}{2} , \forall i \in \{ 1,2 \} , \forall I \in \{ 1 , \cdots , Z \}
\end{multline}
\begin{multline}
    J_{(1)I3} | j_{(1),1}, m_{(1),1}, \cdots , j_{(1),Z},
    m_{(1),Z} , j_{(2),1}, m_{(2),1}, \cdots , j_{(2),Z},
    m_{(2),Z} > \; = \\
     m_{(1),I}  | j_{(1),1}, m_{(1),1}, \cdots , j_{(1),Z},
    m_{(1),Z} , j_{(2),1}, m_{(2),1}, \cdots , j_{(2),Z},
    m_{(2),Z} >  \\
 \forall  m_{(i),I} \in \{ - j_{(i)I} , \cdots , j_{(i)I} \} \, , \, \forall j_{(i)I} \in \frac{\mathbb{N}}{2} , \forall i \in \{ 1,2 \} , \forall I \in \{ 1 , \cdots , Z \}
\end{multline}
\begin{multline}
    \vec{J}_{(2)I}^{2} | j_{(1),1}, m_{(1),1}, \cdots , j_{(1),Z},
    m_{(1),Z} , j_{(2),1}, m_{(2),1}, \cdots , j_{(2),Z},
    m_{(2),Z} > \; = \\
     j_{(2)I} ( j_{(2)I} + 1 ) | j_{(1),1}, m_{(1),1}, \cdots , j_{(1),Z},
    m_{(1),Z} , j_{(2),1}, m_{(2),1}, \cdots , j_{(2),Z},
    m_{(2),Z} > \\
     \forall  m_{(i),I} \in \{ - j_{(i)I} , \cdots , j_{(i)I} \} \, , \, \forall j_{(i)I} \in \frac{\mathbb{N}}{2} , \forall i \in \{ 1,2 \} , \forall I \in \{ 1 , \cdots , Z \}
\end{multline}
\begin{multline}
    J_{(2)I3} | j_{(1),1}, m_{(1),1}, \cdots , j_{(1),Z},
    m_{(1),Z} , j_{(2),1}, m_{(2),1}, \cdots , j_{(2),Z},
    m_{(2),Z} > \; = \\
     m_{(2),I}  | j_{(1),1}, m_{(1),1}, \cdots , j_{(1),Z},
    m_{(1),Z} , j_{(2),1}, m_{(2),1}, \cdots , j_{(2),Z},
    m_{(2),Z} >  \\
      \forall  m_{(i),I} \in \{ - j_{(i)I} , \cdots , j_{(i)I} \} \, , \, \forall j_{(i)I} \in \frac{\mathbb{N}}{2} , \forall i \in \{ 1,2 \} , \forall I \in \{ 1 , \cdots , Z \}
\end{multline}

\bigskip

We can, in an equivalent way, to take into account the set of
Casimir operators $ \{ C_{(1)I},  C_{(2)I} \; I \in \{ 1 , \cdots
, Z \} \} $ where:
\begin{equation}
   C_{(1)I} \; := \; \vec{J}_{(1)I}^{2} +
   \vec{J}_{(2)I}^{2} \; \; I \in \{ 1 , \cdots , Z \}
\end{equation}
\begin{equation}
   C_{(2)I} \; := \; \vec{J}_{(1)I}^{2} -
   \vec{J}_{(2)I}^{2} \; \; I \in \{ 1 , \cdots , Z \}
\end{equation}
Since:
\begin{equation}
    \vec{M}_{I} \cdot \vec{L}_{I} \; = \; 0 \; \; \forall I \in \{
    1 , \cdots , Z \}
\end{equation}
it follows that:
\begin{equation}
     C_{(2)I} \; = \; 0   \; \; \forall I \in \{
    1 , \cdots , Z \}
\end{equation}
so that we can consider only the subspace such that $ j_{(1)I} =
j_{(2)I} \; \forall I \in \{ 1 , \cdots , Z \} $.

Taking into account the equation \ref{eq:rescaled
Laplace-Runge-Lenz operators}, the equation \ref{eq:first angular
momentum operator} and the equation \ref{eq:second angular
momentum operator} it follows that:
\begin{multline}
    H | j_{(1),1}, m_{(1),1}, \cdots , j_{(1),Z},
    m_{(1),Z} , j_{(1),1}, m_{(2),1}, \cdots , j_{(1),Z},
    m_{(2),Z} > \; = \\
    \sum_{I=1}^{Z} E_{j_{(1),I}} | j_{(1),1}, m_{(1),1}, \cdots , j_{(1),Z},
    m_{(1),Z} , j_{(1),1}, m_{(2),1}, \cdots , j_{(1),Z},
    m_{(2),Z} >
\end{multline}
where:
\begin{equation} \label{eq:contribution to the energy level of a single particle}
    E_{j_{(1),I}} \; := \; - \frac{j_{(1),I}^{2}  }{ (2 j_{(1),I} +1 )^{2}
    } \; \; \; j_{(1),I} \in \frac{\mathbb{N}}{2} \,
    , \, I \in \{ 1 , \cdots , Z \}
\end{equation}

\bigskip

Let us now observe that since $ \{ \vec{J}_{(1),I} ,
\vec{J}_{(2),I} \; \; I \in \{ 1 , \cdots , Z \} \} $ are $ 2 Z $
uncoupled angular momentum operators, the equation \ref{eq:angular
momentum is odd under time-reversal} implies that:
\begin{equation}
    T \vec{J}_{(i),I} T^{-1} \; = \;-  \vec{J}_{(i),I} \; \;
    \forall I \in \{ 1 , \cdots , Z \} , \forall i \in \{ 1 , 2 \}
\end{equation}
while the equation \ref{eq:action of time-reversal over
eigenstates of angular momentum} and the equation \ref{eq:action
of squared time-reversal over eigenstates of angular momentum}
imply, respectively, that:
\begin{multline} \label{eq:time-reversal of the 2Z uncoupled angular momenta}
    \overleftarrow{ | j_{(1),1}, m_{(1),1}, \cdots , j_{(1),Z},
    m_{(1),Z} , j_{(2),1}, m_{(2),1}, \cdots , j_{(2),Z},
    m_{(2),Z} > }\; = \\
    i^{2 \sum_{I=1}^{Z} ( m_{(1),I} + m_{(2),I}) } | j_{(1),1}, - m_{(1),1}, \cdots , j_{(1),Z},
    - m_{(1),Z} , j_{(2),1}, -  m_{(2),1}, \cdots , j_{(2),Z},
  -  m_{(2),Z} >
\end{multline}
and:
\begin{multline} \label{eq:squared time-reversal of the 2Z uncoupled angular momenta}
     \overleftarrow{\overleftarrow{| j_{(1),1}, m_{(1),1}, \cdots , j_{(1),Z},
    m_{(1),Z} , j_{(2),1}, m_{(2),1}, \cdots , j_{(2),Z},
    m_{(2),Z} > }}\; = \\
    (-1)^{2 \sum_{I=1}^{Z} ( j_{(1),I} + j_{(2),I}) } | j_{(1),1},  m_{(1),1}, \cdots , j_{(1),Z},
     m_{(1),Z} , j_{(2),1},   m_{(2),1}, \cdots , j_{(2),Z},  m_{(2),Z} >
\end{multline}

\bigskip

Let us now consider the coupling of the quantum angular momenta
operators $ \{ \vec{J}_{(i)I} \; I \in \{ 1 , \cdots , Z \}, i \in
\{ 1 ,2 \} \} $.

Owing to the proposition \ref{prop:coupling of n angular momenta}
it follows that:
\begin{small}
\begin{multline} \label{eq:coupling of 2Z angular momentum}
    | j_{(1)1}, j_{(1)2}, j_{(1)12} , j_{(1)3} , j_{(1)123} , \cdots , j_{(1)1 \cdots
    Z-1} , j_{(1)Z} , j_{(1)}, m_{(1)}, j_{(2)1}, j_{(2)2}, j_{(2)12} , j_{(2)3} , j_{(2)123} , \cdots , j_{(2)1 \cdots
    Z-1} , j_{(2)Z} , j_{(2)}, m_{(2)} > \\ = \; \prod_{i \in \{ 1,2 \} } \sum_{m_{(i)1}=-j_{(i)1}}^{j_{(i)1}}
    \sum_{m_{(i)2}=-j_{(i)2}}^{j_{(i)2}} \cdots
    \sum_{m_{(i)Z}=-j_{(i)Z}}^{j_{(i)Z}}
      C_{j_{(i)1 \cdots Z-1}m_{(i)1 \cdots Z-1}j_{(i)Z}m_{(i)Z}}^{j_{(i)}m_{(i)}}   \cdots   C_{j_{(i)12}m_{(i)12}j_{(i)3}m_{(i)3}}^{j_{(i)123}m_{(i)123}}   C_{j_{(i)1}m_{(i)1}j_{(i)2}m_{(i)2}}^{j_{(i)12}m_{(i)12}}
      \\
    | j_{(1)1}, m_{(1)1} , \cdots , j_{(1)Z}, m_{(1)Z} ,j_{(2)1}, m_{(2)1} , \cdots , j_{(2)Z}, m_{(2)Z}  > \\
      \forall m_{(i)} \in \{ - j_{(i)} , \cdots , j_{(i)} \} , \forall j_{(i)} \in \{| j_{(i)1 \cdots Z-1} - j_{(i)Z} | , \cdots , j_{(i)1 \cdots Z-1}
      +j_{(i)Z} \} \\
 \forall j_{(i)1 \cdots Z-1} \in \{| j_{(i)1 \cdots Z-2} - j_{(i)Z-1} | , \cdots , j_{(i)1 \cdots Z-2}
      +j_{(i)Z-1} \} , \cdots, \forall j_{(i)123} \in \{| j_{(i)12} - j_{(i)3} | , \cdots , j_{(i)12}
      +j_{(i)3} \} , \\
       \forall j_{(i)12} \in \{| j_{(i)1} - j_{(i)2} | , \cdots , j_{(i)1}
      +j_{(i)2} \} , \forall i \in \{ 1,2 \}
\end{multline}
\end{small}
The equation \ref{eq:action of time-reversal over eigenstates of
angular momentum} and the equation \ref{eq:action of squared
time-reversal over eigenstates of angular momentum} imply,
respectively, that:
\begin{small}
\begin{multline}
  \overleftarrow{| j_{(1)1}, j_{(1)2}, j_{(1)12} , j_{(1)3} , j_{(1)123} , \cdots , j_{(1)1 \cdots
    Z-1} , j_{(1)Z} , j_{(1)}, m_{(1)}, j_{(2)1}, j_{(2)2}, j_{(2)12} , j_{(2)3} , j_{(2)123} , \cdots , j_{(2)1 \cdots
    Z-1} , j_{(2)Z} , j_{(2)}, m_{(2)} >} \\
    = \; i^{ 2 ( m_{(1)} + m_{(2)}    ) } \\
    | j_{(1)1}, j_{(1)2}, j_{(1)12} , j_{(1)3} , j_{(1)123} , \cdots , j_{(1)1 \cdots
    Z-1} , j_{(1)Z} , j_{(1)},- m_{(1)}, j_{(2)1}, j_{(2)2}, j_{(2)12} , j_{(2)3} , j_{(2)123} , \cdots , j_{(2)1 \cdots
    Z-1} , j_{(2)Z} , j_{(2)},-  m_{(2)} >
\end{multline}
\begin{multline}
  \overleftarrow{\overleftarrow{| j_{(1)1}, j_{(1)2}, j_{(1)12} , j_{(1)3} , j_{(1)123} , \cdots , j_{(1)1 \cdots
    Z-1} , j_{(1)Z} , j_{(1)}, m_{(1)}, j_{(2)1}, j_{(2)2}, j_{(2)12} , j_{(2)3} , j_{(2)123} , \cdots , j_{(2)1 \cdots
    Z-1} , j_{(2)Z} , j_{(2)}, m_{(2)} >}} \\
    = \; (-1)^{ 2 ( j_{(1)} + j_{(2)}    ) } \\
    | j_{(1)1}, j_{(1)2}, j_{(1)12} , j_{(1)3} , j_{(1)123} , \cdots , j_{(1)1 \cdots
    Z-1} , j_{(1)Z} , j_{(1)}, m_{(1)}, j_{(2)1}, j_{(2)2}, j_{(2)12} , j_{(2)3} , j_{(2)123} , \cdots , j_{(2)1 \cdots
    Z-1} , j_{(2)Z} , j_{(2)}, m_{(2)} >
\end{multline}
\end{small}

\smallskip

Let us now apply the time reversal operator to the equation
\ref{eq:coupling of 2Z angular momentum}; we obtain that:
\begin{small}
\begin{multline}
    \overleftarrow{| j_{(1)1}, j_{(1)2}, j_{(1)12} , j_{(1)3} , j_{(1)123} , \cdots , j_{(1)1 \cdots
    Z-1} , j_{(1)Z} , j_{(1)}, m_{(1)}, j_{(2)1}, j_{(2)2}, j_{(2)12} , j_{(2)3} , j_{(2)123} , \cdots , j_{(2)1 \cdots
    Z-1} , j_{(2)Z} , j_{(2)}, m_{(2)} >} \\ = \; \prod_{i \in \{ 1,2 \} } \sum_{m_{(i)1}=-j_{(i)1}}^{j_{(i)1}}
    \sum_{m_{(i)2}=-j_{(i)2}}^{j_{(i)2}} \cdots
    \sum_{m_{(i)Z}=-j_{(i)Z}}^{j_{(i)Z}}
     \overline{ C_{j_{(i)1 \cdots Z-1}m_{(i)1 \cdots Z-1}j_{(i)Z}m_{(i)Z}}^{j_{(i)}m_{(i)}}   \cdots   C_{j_{(i)12}m_{(i)12}j_{(i)3}m_{(i)3}}^{j_{(i)123}m_{(i)123}}   C_{j_{(i)1}m_{(i)1}j_{(i)2}m_{(i)2}}^{j_{(i)12}m_{(i)12}}}
      \\
    \overleftarrow{| j_{(1)1}, m_{(1)1} , \cdots , j_{(1)Z}, m_{(1)Z} ,j_{(2)1}, m_{(2)1} , \cdots , j_{(2)Z}, m_{(2)Z}  >} \\
      \forall m_{(i)} \in \{ - j_{(i)} , \cdots , j_{(i)} \} , \forall j_{(i)} \in \{| j_{(i)1 \cdots Z-1} - j_{(i)Z} | , \cdots , j_{(i)1 \cdots Z-1}
      +j_{(i)Z} \} \\
 \forall j_{(i)1 \cdots Z-1} \in \{| j_{(i)1 \cdots Z-2} - j_{(i)Z-1} | , \cdots , j_{(i)1 \cdots Z-2}
      +j_{(i)Z-1} \} , \cdots, \forall j_{(i)123} \in \{| j_{(i)12} - j_{(i)3} | , \cdots , j_{(i)12}
      +j_{(i)3} \} , \\
       \forall j_{(i)12} \in \{| j_{(i)1} - j_{(i)2} | , \cdots , j_{(i)1}
      +j_{(i)2} \} , \forall i \in \{ 1,2 \}
\end{multline}
\end{small}
\begin{small}
\begin{multline}
    \overleftarrow{| j_{(1)1}, j_{(1)2}, j_{(1)12} , j_{(1)3} , j_{(1)123} , \cdots , j_{(1)1 \cdots
    Z-1} , j_{(1)Z} , j_{(1)}, m_{(1)}, j_{(2)1}, j_{(2)2}, j_{(2)12} , j_{(2)3} , j_{(2)123} , \cdots , j_{(2)1 \cdots
    Z-1} , j_{(2)Z} , j_{(2)}, m_{(2)} >} \\ = \; \prod_{i \in \{ 1,2 \} } \sum_{m_{(i)1}=-j_{(i)1}}^{j_{(i)1}}
    \sum_{m_{(i)2}=-j_{(i)2}}^{j_{(i)2}} \cdots
    \sum_{m_{(i)Z}=-j_{(i)Z}}^{j_{(i)Z}}
      C_{j_{(i)1 \cdots Z-1}m_{(i)1 \cdots Z-1}j_{(i)Z}m_{(i)Z}}^{j_{(i)}m_{(i)}}   \cdots   C_{j_{(i)12}m_{(i)12}j_{(i)3}m_{(i)3}}^{j_{(i)123}m_{(i)123}}   C_{j_{(i)1}m_{(i)1}j_{(i)2}m_{(i)2}}^{j_{(i)12}m_{(i)12}}
      \\
    \overleftarrow{| j_{(1)1},  m_{(1)1} , \cdots , j_{(1)Z},  m_{(1)Z} ,j_{(2)1},  m_{(2)1} , \cdots , j_{(2)Z},  m_{(2)Z}  > }\\
      \forall m_{(i)} \in \{ - j_{(i)} , \cdots , j_{(i)} \} , \forall j_{(i)} \in \{| j_{(i)1 \cdots Z-1} - j_{(i)Z} | , \cdots , j_{(i)1 \cdots Z-1}
      +j_{(i)Z} \} \\
 \forall j_{(i)1 \cdots Z-1} \in \{| j_{(i)1 \cdots Z-2} - j_{(i)Z-1} | , \cdots , j_{(i)1 \cdots Z-2}
      +j_{(i)Z-1} \} , \cdots, \forall j_{(i)123} \in \{| j_{(i)12} - j_{(i)3} | , \cdots , j_{(i)12}
      +j_{(i)3} \} , \\
       \forall j_{(i)12} \in \{| j_{(i)1} - j_{(i)2} | , \cdots , j_{(i)1}
      +j_{(i)2} \} , \forall i \in \{ 1,2 \}
\end{multline}
\end{small}
where we have used the anti-linearity of T and the reality of the
Clebsch-Gordan coefficients. Using the equation
\ref{eq:time-reversal of the 2Z uncoupled angular momenta} it
follows that:
\begin{small}
\begin{multline} \label{eq:expanded time-reversal of 2Z uncoupled angular momenta}
    \overleftarrow{| j_{(1)1}, j_{(1)2}, j_{(1)12} , j_{(1)3} , j_{(1)123} , \cdots , j_{(1)1 \cdots
    Z-1} , j_{(1)Z} , j_{(1)}, m_{(1)}, j_{(2)1}, j_{(2)2}, j_{(2)12} , j_{(2)3} , j_{(2)123} , \cdots , j_{(2)1 \cdots
    Z-1} , j_{(2)Z} , j_{(2)}, m_{(2)} >} \\ = \; \prod_{i \in \{ 1,2 \} } \sum_{m_{(i)1}=-j_{(i)1}}^{j_{(i)1}}
    \sum_{m_{(i)2}=-j_{(i)2}}^{j_{(i)2}} \cdots
    \sum_{m_{(i)Z}=-j_{(i)Z}}^{j_{(i)Z}}
      C_{j_{(i)1 \cdots Z-1}m_{(i)1 \cdots Z-1}j_{(i)Z}m_{(i)Z}}^{j_{(i)}m_{(i)}}   \cdots   C_{j_{(i)12}m_{(i)12}j_{(i)3}m_{(i)3}}^{j_{(i)123}m_{(i)123}}   C_{j_{(i)1}m_{(i)1}j_{(i)2}m_{(i)2}}^{j_{(i)12}m_{(i)12}}
      \\
   i^{2 \sum_{I=1}^{Z} ( m_{(1),I} + m_{(2),I}) }  | j_{(1)1}, - m_{(1)1} , \cdots , j_{(1)Z}, - m_{(1)Z} ,j_{(2)1}, - m_{(2)1} , \cdots , j_{(2)Z}, - m_{(2)Z}  > \\
      \forall m_{(i)} \in \{ - j_{(i)} , \cdots , j_{(i)} \} , \forall j_{(i)} \in \{| j_{(i)1 \cdots Z-1} - j_{(i)Z} | , \cdots , j_{(i)1 \cdots Z-1}
      +j_{(i)Z} \} \\
 \forall j_{(i)1 \cdots Z-1} \in \{| j_{(i)1 \cdots Z-2} - j_{(i)Z-1} | , \cdots , j_{(i)1 \cdots Z-2}
      +j_{(i)Z-1} \} , \cdots, \forall j_{(i)123} \in \{| j_{(i)12} - j_{(i)3} | , \cdots , j_{(i)12}
      +j_{(i)3} \} , \\
       \forall j_{(i)12} \in \{| j_{(i)1} - j_{(i)2} | , \cdots , j_{(i)1}
      +j_{(i)2} \} , \forall i \in \{ 1,2 \}
\end{multline}
\end{small}
Comparing  the equation \ref{eq:time-reversal of the 2Z uncoupled
angular momenta} and the equation \ref{eq:expanded time-reversal
of 2Z uncoupled angular momenta} it follows that:
\begin{small}
\begin{multline}
    | j_{(1)1}, j_{(1)2}, j_{(1)12} , j_{(1)3} , j_{(1)123} , \cdots , j_{(1)1 \cdots
    Z-1} , j_{(1)Z} , j_{(1)}, - m_{(1)}, j_{(2)1}, j_{(2)2}, j_{(2)12} , j_{(2)3} , j_{(2)123} , \cdots , j_{(2)1 \cdots
    Z-1} , j_{(2)Z} , j_{(2)}, -  m_{(2)} > \\ = \; \prod_{i \in \{ 1,2 \} } \sum_{m_{(i)1}=-j_{(i)1}}^{j_{(i)1}}
    \sum_{m_{(i)2}=-j_{(i)2}}^{j_{(i)2}} \cdots
    \sum_{m_{(i)Z}=-j_{(i)Z}}^{j_{(i)Z}}
      C_{j_{(i)1 \cdots Z-1}m_{(i)1 \cdots Z-1}j_{(i)Z}m_{(i)Z}}^{j_{(i)}m_{(i)}}   \cdots   C_{j_{(i)12}m_{(i)12}j_{(i)3}m_{(i)3}}^{j_{(i)123}m_{(i)123}}   C_{j_{(i)1}m_{(i)1}j_{(i)2}m_{(i)2}}^{j_{(i)12}m_{(i)12}}
      \\
   i^{2 [ m_{(1)} + m_{(2)} - \sum_{I=1}^{Z} ( m_{(1),I} + m_{(2),I}) ]}  | j_{(1)1}, - m_{(1)1} , \cdots , j_{(1)Z}, - m_{(1)Z} ,j_{(2)1}, - m_{(2)1} , \cdots , j_{(2)Z}, - m_{(2)Z}  > \\
      \forall m_{(i)} \in \{ - j_{(i)} , \cdots , j_{(i)} \} , \forall j_{(i)} \in \{| j_{(i)1 \cdots Z-1} - j_{(i)Z} | , \cdots , j_{(i)1 \cdots Z-1}
      +j_{(i)Z} \} \\
 \forall j_{(i)1 \cdots Z-1} \in \{| j_{(i)1 \cdots Z-2} - j_{(i)Z-1} | , \cdots , j_{(i)1 \cdots Z-2}
      +j_{(i)Z-1} \} , \cdots, \forall j_{(i)123} \in \{| j_{(i)12} - j_{(i)3} | , \cdots , j_{(i)12}
      +j_{(i)3} \} , \\
       \forall j_{(i)12} \in \{| j_{(i)1} - j_{(i)2} | , \cdots , j_{(i)1}
      +j_{(i)2} \} , \forall i \in \{ 1,2 \}
\end{multline}
\end{small}
that, using the proposition \ref{prop:basic property of the
Clebsch-Gordan coefficient}, implies that:
\begin{proposition}
\end{proposition}
\begin{small}
\begin{multline}
    | j_{(1)1}, j_{(1)2}, j_{(1)12} , j_{(1)3} , j_{(1)123} , \cdots , j_{(1)1 \cdots
    Z-1} , j_{(1)Z} , j_{(1)}, - m_{(1)}, j_{(2)1}, j_{(2)2}, j_{(2)12} , j_{(2)3} , j_{(2)123} , \cdots , j_{(2)1 \cdots
    Z-1} , j_{(2)Z} , j_{(2)}, -  m_{(2)} > \\ = \; \prod_{i \in \{ 1,2 \} } \sum_{m_{(i)1}=-j_{(i)1}}^{j_{(i)1}}
    \sum_{m_{(i)2}=-j_{(i)2}}^{j_{(i)2}} \cdots
    \sum_{m_{(i)Z}=-j_{(i)Z}}^{j_{(i)Z}}
      C_{j_{(i)1 \cdots Z-1}m_{(i)1 \cdots Z-1}j_{(i)Z}m_{(i)Z}}^{j_{(i)}m_{(i)}}   \cdots   C_{j_{(i)12}m_{(i)12}j_{(i)3}m_{(i)3}}^{j_{(i)123}m_{(i)123}}   C_{j_{(i)1}m_{(i)1}j_{(i)2}m_{(i)2}}^{j_{(i)12}m_{(i)12}}
      \\
     | j_{(1)1}, - m_{(1)1} , \cdots , j_{(1)Z}, - m_{(1)Z} ,j_{(2)1}, - m_{(2)1} , \cdots , j_{(2)Z}, - m_{(2)Z}  > \\
      \forall m_{(i)} \in \{ - j_{(i)} , \cdots , j_{(i)} \} , \forall j_{(i)} \in \{| j_{(i)1 \cdots Z-1} - j_{(i)Z} | , \cdots , j_{(i)1 \cdots Z-1}
      +j_{(i)Z} \} \\
 \forall j_{(i)1 \cdots Z-1} \in \{| j_{(i)1 \cdots Z-2} - j_{(i)Z-1} | , \cdots , j_{(i)1 \cdots Z-2}
      +j_{(i)Z-1} \} , \cdots, \forall j_{(i)123} \in \{| j_{(i)12} - j_{(i)3} | , \cdots , j_{(i)12}
      +j_{(i)3} \} , \\
       \forall j_{(i)12} \in \{| j_{(i)1} - j_{(i)2} | , \cdots , j_{(i)1}
      +j_{(i)2} \} , \forall i \in \{ 1,2 \}
\end{multline}
\end{small}

\smallskip

Let us now apply the squared time reversal operator to the
equation \ref{eq:coupling of 2Z angular momentum}; we obtain that:
\begin{small}
\begin{multline}
    \overleftarrow{\overleftarrow{| j_{(1)1}, j_{(1)2}, j_{(1)12} , j_{(1)3} , j_{(1)123} , \cdots , j_{(1)1 \cdots
    Z-1} , j_{(1)Z} , j_{(1)}, m_{(1)}, j_{(2)1}, j_{(2)2}, j_{(2)12} , j_{(2)3} , j_{(2)123} , \cdots , j_{(2)1 \cdots
    Z-1} , j_{(2)Z} , j_{(2)}, m_{(2)} > }} \\ = \; \prod_{i \in \{ 1,2 \} } \sum_{m_{(i)1}=-j_{(i)1}}^{j_{(i)1}}
    \sum_{m_{(i)2}=-j_{(i)2}}^{j_{(i)2}} \cdots
    \sum_{m_{(i)Z}=-j_{(i)Z}}^{j_{(i)Z}}
      C_{j_{(i)1 \cdots Z-1}m_{(i)1 \cdots Z-1}j_{(i)Z}m_{(i)Z}}^{j_{(i)}m_{(i)}}   \cdots   C_{j_{(i)12}m_{(i)12}j_{(i)3}m_{(i)3}}^{j_{(i)123}m_{(i)123}}   C_{j_{(i)1}m_{(i)1}j_{(i)2}m_{(i)2}}^{j_{(i)12}m_{(i)12}}
      \\
    \overleftarrow{\overleftarrow{| j_{(1)1}, m_{(1)1} , \cdots , j_{(1)Z}, m_{(1)Z} ,j_{(2)1}, m_{(2)1} , \cdots , j_{(2)Z}, m_{(2)Z}  > }}\\
      \forall m_{(i)} \in \{ - j_{(i)} , \cdots , j_{(i)} \} , \forall j_{(i)} \in \{| j_{(i)1 \cdots Z-1} - j_{(i)Z} | , \cdots , j_{(i)1 \cdots Z-1}
      +j_{(i)Z} \} \\
 \forall j_{(i)1 \cdots Z-1} \in \{| j_{(i)1 \cdots Z-2} - j_{(i)Z-1} | , \cdots , j_{(i)1 \cdots Z-2}
      +j_{(i)Z-1} \} , \cdots, \forall j_{(i)123} \in \{| j_{(i)12} - j_{(i)3} | , \cdots , j_{(i)12}
      +j_{(i)3} \} , \\
       \forall j_{(i)12} \in \{| j_{(i)1} - j_{(i)2} | , \cdots , j_{(i)1}
      +j_{(i)2} \} , \forall i \in \{ 1,2 \}
\end{multline}
\end{small}
where we have used the linearity of $ T^{2} $.

Using the equation \ref{eq:squared time-reversal of the 2Z
uncoupled angular momenta} it follows that:
\begin{small}
\begin{multline} \label{eq:expanded squared time-reversal of 2Z uncoupled angular momenta}
    \overleftarrow{\overleftarrow{| j_{(1)1}, j_{(1)2}, j_{(1)12} , j_{(1)3} , j_{(1)123} , \cdots , j_{(1)1 \cdots
    Z-1} , j_{(1)Z} , j_{(1)}, m_{(1)}, j_{(2)1}, j_{(2)2}, j_{(2)12} , j_{(2)3} , j_{(2)123} , \cdots , j_{(2)1 \cdots
    Z-1} , j_{(2)Z} , j_{(2)}, m_{(2)} > }} \\ = \; \prod_{i \in \{ 1,2 \} } \sum_{m_{(i)1}=-j_{(i)1}}^{j_{(i)1}}
    \sum_{m_{(i)2}=-j_{(i)2}}^{j_{(i)2}} \cdots
    \sum_{m_{(i)Z}=-j_{(i)Z}}^{j_{(i)Z}}
      C_{j_{(i)1 \cdots Z-1}m_{(i)1 \cdots Z-1}j_{(i)Z}m_{(i)Z}}^{j_{(i)}m_{(i)}}   \cdots   C_{j_{(i)12}m_{(i)12}j_{(i)3}m_{(i)3}}^{j_{(i)123}m_{(i)123}}   C_{j_{(i)1}m_{(i)1}j_{(i)2}m_{(i)2}}^{j_{(i)12}m_{(i)12}}
      \\
    ( -1)^{2 \sum_{I=1}^{Z} ( j_{(1)I} +  j_{(2)I} )  } | j_{(1)1}, m_{(1)1} , \cdots , j_{(1)Z}, m_{(1)Z} ,j_{(2)1}, m_{(2)1} , \cdots , j_{(2)Z}, m_{(2)Z}  > \\
      \forall m_{(i)} \in \{ - j_{(i)} , \cdots , j_{(i)} \} , \forall j_{(i)} \in \{| j_{(i)1 \cdots Z-1} - j_{(i)Z} | , \cdots , j_{(i)1 \cdots Z-1}
      +j_{(i)Z} \} \\
 \forall j_{(i)1 \cdots Z-1} \in \{| j_{(i)1 \cdots Z-2} - j_{(i)Z-1} | , \cdots , j_{(i)1 \cdots Z-2}
      +j_{(i)Z-1} \} , \cdots, \forall j_{(i)123} \in \{| j_{(i)12} - j_{(i)3} | , \cdots , j_{(i)12}
      +j_{(i)3} \} , \\
       \forall j_{(i)12} \in \{| j_{(i)1} - j_{(i)2} | , \cdots , j_{(i)1}
      +j_{(i)2} \} , \forall i \in \{ 1,2 \}
\end{multline}
\end{small}

Comparing  the equation \ref{eq:squared time-reversal of the 2Z
uncoupled angular momenta} and the equation \ref{eq:expanded
squared time-reversal of 2Z uncoupled angular momenta} it follows
that:
\begin{proposition} \label{prop:consequence of squared time reversal of 2Z angular momenta}
\end{proposition}
\begin{small}
\begin{multline}
    | j_{(1)1}, j_{(1)2}, j_{(1)12} , j_{(1)3} , j_{(1)123} , \cdots , j_{(1)1 \cdots
    Z-1} , j_{(1)Z} , j_{(1)}, m_{(1)}, j_{(2)1}, j_{(2)2}, j_{(2)12} , j_{(2)3} , j_{(2)123} , \cdots , j_{(2)1 \cdots
    Z-1} , j_{(2)Z} , j_{(2)}, m_{(2)} > \\ = \; \prod_{i \in \{ 1,2 \} } \sum_{m_{(i)1}=-j_{(i)1}}^{j_{(i)1}}
    \sum_{m_{(i)2}=-j_{(i)2}}^{j_{(i)2}} \cdots
    \sum_{m_{(i)Z}=-j_{(i)Z}}^{j_{(i)Z}}
      C_{j_{(i)1 \cdots Z-1}m_{(i)1 \cdots Z-1}j_{(i)Z}m_{(i)Z}}^{j_{(i)}m_{(i)}}   \cdots   C_{j_{(i)12}m_{(i)12}j_{(i)3}m_{(i)3}}^{j_{(i)123}m_{(i)123}}   C_{j_{(i)1}m_{(i)1}j_{(i)2}m_{(i)2}}^{j_{(i)12}m_{(i)12}}
      \\
   ( -1)^{2 (\sum_{I=1}^{Z} ( j_{(1)I} +  j_{(2)I} ) - j_{(1)}-j_{2}) }  | j_{(1)1}, m_{(1)1} , \cdots , j_{(1)Z}, m_{(1)Z} ,j_{(2)1}, m_{(2)1} , \cdots , j_{(2)Z}, m_{(2)Z}  > \\
      \forall m_{(i)} \in \{ - j_{(i)} , \cdots , j_{(i)} \} , \forall j_{(i)} \in \{| j_{(i)1 \cdots Z-1} - j_{(i)Z} | , \cdots , j_{(i)1 \cdots Z-1}
      +j_{(i)Z} \} \\
 \forall j_{(i)1 \cdots Z-1} \in \{| j_{(i)1 \cdots Z-2} - j_{(i)Z-1} | , \cdots , j_{(i)1 \cdots Z-2}
      +j_{(i)Z-1} \} , \cdots, \forall j_{(i)123} \in \{| j_{(i)12} - j_{(i)3} | , \cdots , j_{(i)12}
      +j_{(i)3} \} , \\
       \forall j_{(i)12} \in \{| j_{(i)1} - j_{(i)2} | , \cdots , j_{(i)1}
      +j_{(i)2} \} , \forall i \in \{ 1,2 \}
\end{multline}
\end{small}

Let us now remark that, exactly as in the analogous computation
performed in the section \ref{sec:Considerations about the double
time-reversal superselection rule in Nonrelativistic Quantum
Mechanics}, the equation \ref{eq:coupling of 2Z angular momentum}
and the proposition \ref{prop:consequence of squared time reversal
of 2Z angular momenta} are consistent if and only if:
\begin{multline} \label{eq:consistency condition for 2Z angular momenta}
    (-1)^{2 (\sum_{I=1}^{Z} ( j_{(1)I} +  j_{(2)I} ) -
    j_{(1)}-j_{2})
    } \; = \, 1 \\
     \forall j_{(1)} \in \{ \mathcal{J}_{min} ( j_{(1)1} , \cdots
    , j_{(1)Z} ) , \cdots , \mathcal{J}_{max} ( j_{(1)1} , \cdots
    , j_{(1)Z} ) \}   ,
       \forall j_{(2)} \in \{ \mathcal{J}_{min} ( j_{(2)1} , \cdots
    , j_{(2)Z} ) , \cdots , \mathcal{J}_{max} ( j_{(2)1} , \cdots
    , j_{(2)Z} ) \} , \\
      \forall j_{iI} \in \frac{\mathbb{N}}{2}
 ,  \forall i \in \{ 1,2 \} \, , \, \forall I \in \{ 1 , \cdots , Z \}
\end{multline}

Again the equation \ref{eq:consistency condition for 2Z angular
momenta} is, anyway, a trivial consequence of the proposition
\ref{prop:univalence of coupled angular momenta}.

\bigskip

Let us at last analyze the degeneracy of the energy levels.

We have obviously that:
\begin{multline}
    H | j_{(1)1}, m_{(1)1} , \cdots , j_{(1)Z}, m_{(1)Z} ,j_{(2)1}:=j_{(1)1}, m_{(2)1} , \cdots , j_{(2)Z}:=j_{(1)Z}, m_{(2)Z}
    > \; = \\
      E_{j_{(1),1}, \cdots , j_{(1),Z} }  | j_{(1)1}, m_{(1)1} , \cdots , j_{(1)Z}, m_{(1)Z} ,j_{(2)1}:=j_{(1)1}, m_{(2)1} , \cdots , j_{(2)Z}:=j_{(1)Z}, m_{(2)Z} >
\end{multline}
where:
\begin{equation} \label{eq:energy levels of 2Z bosons in a Keplerian field}
    E_{j_{(1),1}, \cdots , j_{(1),Z} } \; := \; \sum_{I=1}^{Z}  E_{j_{(1),I}}
\end{equation}
so that:
\begin{equation} \label{eq:degeneration of the energy levels for 2Z bosons}
    degeneration(E_{j_{(1),1}, \cdots , j_{(1),Z} } ) \; = \;
    2 \prod_{I=1}^{Z}  ( 2 j_{(1),I} +1 )
\end{equation}
where we recall that $ E_{j_{(1),I}} $ is defined in the equation
\ref{eq:contribution to the energy level of a single particle} and
when the factor 2 in the right hand side of the equation
\ref{eq:degeneration of the energy levels for 2Z bosons} is owed
to the fact that  $ j_{(2),I} $ is equal to $ j_{(1)I} $ but, in
general $ m_{(2),I} $ is not equal to $ m_{(1),I} $.

Such a degeneration cannot, anyway, be inferred from the theorem
\ref{th:Wigner's Theorem about Kramers degeneracy} since the
angular momentum of the involved system is only the \emph{orbital}
one $ \vec{L} $ that has always univalence equal to + 1.

\smallskip

One could, at this point, think to adapt the proof of the theorem
\ref{th:Wigner's Theorem about Kramers degeneracy} observing that
given $ j_{(1),1} , \cdots , j_{(1),Z}, j_{(2),1} , \cdots ,
j_{(2),Z} \in \frac{\mathbb{N}}{2} $ and such that:
\begin{multline}
  \overleftarrow{\overleftarrow{| j_{(1)1}, m_{(1)1} , \cdots , j_{(1)Z}, m_{(1)Z} ,j_{(2)1}:=j_{(1)1}, m_{(2)1} , \cdots , j_{(2)Z}:=j_{(1)Z}, m_{(2)Z}
    >}} \; = \\
     -  | j_{(1)1}, m_{(1)1} , \cdots , j_{(1)Z}, m_{(1)Z} ,j_{(2)1}:=j_{(1)1}, m_{(2)1} , \cdots , j_{(2)Z}:=j_{(1)Z}, m_{(2)Z}
    >
\end{multline}
the proposition \ref{prop:basic property of states odd under
squared time-reversal} would imply that:
\begin{tiny}
\begin{multline}
  < j_{(1)1}, m_{(1)1} , \cdots , j_{(1)Z}, m_{(1)Z} ,j_{(2)1}:=j_{(1)1}, m_{(2)1} , \cdots , j_{(2)Z}:=j_{(1)Z}, m_{(2)Z}
    \overleftarrow{| j_{(1)1}, m_{(1)1} , \cdots , j_{(1)Z}, m_{(1)Z} ,j_{(2)1}:=j_{(1)1}, m_{(2)1} , \cdots , j_{(2)Z}:=j_{(1)Z}, m_{(2)Z}
    >} \\
     = \; 0
\end{multline}
\end{tiny}
that, combined with the equation \ref{eq:symmetry of the
hamiltonian} assuring that T leaves invariant any eigenspace of H,
would allow to infer that each energy level is at least doubly
degenerate.

Such an argument is not, anyway, correct since:
\begin{multline}
    j_{(2)I} = j_{(1)I} \, \, \forall I \in \{ 1 , \cdots , Z \}
    \; \Rightarrow \\
     [ ( -1)^{2 \sum_{I=1}^{Z} ( j_{(1)I} + j_{(2)I} )
    } \, = \, ( -1)^{4 \sum_{I=1}^{Z} j_{(1)I}} = 1 \; \; \forall
    j_{(1)I} \in \frac{N}{2} , \forall I \in \{ 1 , \cdots , Z \} ]
\end{multline}
and hence:
\begin{multline}
  \overleftarrow{\overleftarrow{| j_{(1)1}, m_{(1)1} , \cdots , j_{(1)Z}, m_{(1)Z} ,j_{(2)1}:=j_{(1)1}, m_{(2)1} , \cdots , j_{(2)Z}:=j_{(1)Z}, m_{(2)Z}
    >}} \; = \\
     +| j_{(1)1}, m_{(1)1} , \cdots , j_{(1)Z}, m_{(1)Z} ,j_{(2)1}:=j_{(1)1}, m_{(2)1} , \cdots , j_{(2)Z}:=j_{(1)Z}, m_{(2)Z} >
\end{multline}

\bigskip

\begin{example} \label{ex:Z fermions in a Keplerian field}
\end{example}
Let us now consider a situation identical to the one discussed in
the example \ref{ex:Z bosons in a Keplerian field} but for the
fact that the $ Z \in \mathbb{N} : n > 2 $ involved particles are,
this time, fermions of spin 1/2.

Introduced the total spin:
\begin{equation}
    \vec{S} \; := \; \sum_{I=1}^{Z} \vec{S}_{I}
\end{equation}
let us remark, first of all, that obviously:
\begin{equation}
   [ H ,  \vec{S}_{(I)} ] \; = \; [ H , S_{(I)3} ] \; = \; [ H ,  \vec{S} ] \; = \; [ H , S_{3} ] \; = \;
   0 \; \; \forall I \in \{ 1 , \cdots , Z \}
\end{equation}
Clearly:
\begin{multline}
    H | j_{(1)1}, m_{(1)1} , \cdots , j_{(1)Z}, m_{(1)Z} ,j_{(2)1}:=j_{(1)1}, m_{(2)1} , \cdots , j_{(2)Z}:=j_{(1)Z},
    m_{(2)Z},
    m_{(s)1} , \cdots , m_{(s)Z}> \; = \\
      E_{j_{(1),1}, \cdots , j_{(1),Z}} | j_{(1)1}, m_{(1)1} , \cdots , j_{(1)Z}, m_{(1)Z} ,j_{(2)1}:=j_{(1)1}, m_{(2)1} , \cdots , j_{(2)Z}:=j_{(1)Z}, m_{(2)Z},
    m_{(s)1} , \cdots , m_{(s)Z}>
\end{multline}
where $ E_{j_{(1),1}, \cdots , j_{(1),Z} } $ is given again by the
equation \ref{eq:contribution to the energy level of a single
particle} and by the equation \ref{eq:energy levels of 2Z bosons
in a Keplerian field}.

The degeneration of the energy levels, anyway, is this time given
by
\begin{equation} \label{eq:degeneration of the energy levels for 2Z fermions}
    degeneration(E_{j_{(1),1}, \cdots , j_{(1),Z} } ) \; = \;
    2Z +  2 \prod_{I=1}^{Z}  ( 2 j_{(1),I} +1 )
\end{equation}
owing to the 2 possible values that each $ m_{(s)I} $ can take.

Let us now observe that:
\begin{multline}
  \overleftarrow{\overleftarrow{| j_{(1)1}, m_{(1)1} , \cdots , j_{(1)Z}, m_{(1)Z} ,j_{(2)1}:=j_{(1)1}, m_{(2)1} , \cdots , j_{(2)Z}:=j_{(1)Z}, m_{(2)Z}
   , m_{(s)1} , \cdots , m_{(s)Z}>}} \; = \\
     ( -1)^{4 \sum_{I=1}^{Z} j_{(1)I }} ( -1 )^{Z} | j_{(1)1}, m_{(1)1} , \cdots , j_{(1)Z}, m_{(1)Z} ,j_{(2)1}:=j_{(1)1}, m_{(2)1} , \cdots , j_{(2)Z}:=j_{(1)Z},
     m_{(2)Z},
    m_{(s)1} , \cdots , m_{(s)Z}>
\end{multline}
and hence:
\begin{multline}
  \overleftarrow{\overleftarrow{| j_{(1)1}, m_{(1)1} , \cdots , j_{(1)Z}, m_{(1)Z} ,j_{(2)1}:=j_{(1)1}, m_{(2)1} , \cdots , j_{(2)Z}:=j_{(1)Z}, m_{(2)Z}
   , m_{(s)1} , \cdots , m_{(s)Z}>}} \; = \\
\left\{%
\begin{array}{ll}
    +  | j_{(1)1}, m_{(1)1} , \cdots , j_{(1)Z}, m_{(1)Z} ,j_{(2)1}:=j_{(1)1}, m_{(2)1} , \cdots , j_{(2)Z}:=j_{(1)Z},
    m_{(2)Z},  m_{(s)1} , \cdots , m_{(s)Z}>, & \hbox{if $ Z \in \mathbb{E}$;} \\
    - | j_{(1)1}, m_{(1)1} , \cdots , j_{(1)Z}, m_{(1)Z} ,j_{(2)1}:=j_{(1)1}, m_{(2)1} , \cdots , j_{(2)Z}:=j_{(1)Z},
    m_{(2)Z}, m_{(s)1} , \cdots , m_{(s)Z}>, & \hbox{if $ Z \in \mathbb{O}$.} \\
\end{array}%
\right.
\end{multline}

If the number Z of involved fermions is even no information about
the degeneracy of the energy levels can, consequentially, be
inferred by the \emph{double-time-reversal} symmetry of the
hamiltonian.

Let us suppose, contrary, that $ Z \in \mathbb{O} $.

Then, posed:
\begin{equation}
    | \psi > \; := \; | j_{(1)1}, m_{(1)1} , \cdots , j_{(1)Z}, m_{(1)Z} , j_{(2)1}:=j_{(1)1}, m_{(2)1} , \cdots , j_{(2)Z}:=j_{(1)Z},
  m_{(2)Z} ,  m_{(s)1} , \cdots , m_{(s)Z} >
\end{equation}
the proposition \ref{prop:basic property of states odd under
squared time-reversal} implies that:
\begin{equation}
    < \psi\overleftarrow{ | \psi >} \; = \; 0
\end{equation}
that, combined with the equation \ref{eq:symmetry of the
hamiltonian}  assuring that T leaves invariant any eigenspace of
H, allows to infer that each energy level is at least doubly
degenerate.

Let us remark, anyway that such an inference is not an application
of the theorem \ref{th:Wigner's Theorem about Kramers degeneracy}
since it hasn't been obtained taking into account the univalence
of the angular momentum $ \vec{L} + \vec{S} $.

\newpage

\appendix
\section{Some useful algebraic properties of the set of the half-integer numbers} \label{sec:Some useful algebraic properties}

Obviously $  ( \frac{\mathbb{N}}{2} , + ) $ is a commutative
group.

Introduced:
\begin{itemize}
    \item  the \emph{set of the even numbers}:
\begin{equation}
     \mathbb{E} \; := \; \{ 2 n , n \in \mathbb{N} \}
\end{equation}
    \item the \emph{set of the odd numbers}:
\begin{equation}
     \mathbb{O} \; := \; \{ 2 n+1 , n \in \mathbb{N} \}
\end{equation}
    \item the \emph{set of the half-odd numbers}:
\begin{equation}
     \mathbb{H} \; := \; \frac{\mathbb{O}}{2}
\end{equation}
\end{itemize}
one has clearly that:
\begin{proposition} \label{prop:algebra of the integers union the half-integral}
\end{proposition}
\begin{equation}
    \frac{\mathbb{N}}{2} \; = \; \mathbb{N} \cup \mathbb{H}
\end{equation}
\begin{equation}
    \mathbb{N} \cap \mathbb{H} \; = \; \emptyset
\end{equation}
\begin{equation}
    x+ y \in \mathbb{E} \; \; \forall x,y \in \mathbb{E}
\end{equation}
\begin{equation}
     x+ y \in \mathbb{O} \; \; \forall x \in \mathbb{E} , \forall
     y \in \mathbb{O}
\end{equation}
\begin{equation}
     x+ y \in \mathbb{E} \; \; \forall x \in \mathbb{O} , \forall
     y \in \mathbb{O}
\end{equation}
\begin{equation}
     x+ y \in \mathbb{H} \; \; \forall x \in \mathbb{N} , \forall
     y \in \mathbb{H}
\end{equation}
\begin{equation}
   x+ y \in \mathbb{N} \; \; \forall x,y \in \mathbb{H}
\end{equation}
\begin{equation}
    | x - y | \in \mathbb{N} \; \; \forall x,y \in \mathbb{H}
\end{equation}
\begin{equation}
   | x - y | \in \mathbb{H} \; \; \forall x \in \mathbb{N}, \forall
   y \in \mathbb{H}
\end{equation}
\newpage
\section{Some meta-textual convention}
I would like, first of all, to clarify that my use, in this as in
any of my previous papers, of the first person plural has not to
be considered as an act of arrogance (id est as a sort of
\emph{pluralis maiestatis)} but as its opposite: as an act of
modesty performed in order to include the reader in the
dissertation.

\smallskip

Then I would like to remark that the dates reported in the
bibliography are those of the edition of the books that I have
used.

This might lead some reader to misleading conclusions concerning
the historical  development of ideas: for instance the  $ 3^{th} $
section "Perpetual motion" of the $ 6^{th} $ chapter "Statistical
Entropy" of Oliver Penrose's book \cite{Penrose-05} historically
predates  Charles Bennett's later exorcism of Maxwell's demon
based on the observation that, assuming the Landauer's Principle
stating that a recursive function  can be computed without
increasing the entropy of the Universe if and only if it is
injective, the erasure of the demon's memory cannot be performed
without increasing the entropy of the Universe and that it is
exactly the contribution to the thermodynamical balance of the
erasure of demon's memory that preserves from the violation of the
Second Principle of Thermodynamics.

\smallskip
Last but not least I advise the reader that in this paper it has
been used a Unit System in which $ \hbar = 1 $.

\newpage
\section{Acknowledgements}
I would like to thank strongly Vittorio de Alfaro for his
friendship and his moral support, without which I would have
already given up.

Then I would like to thank strongly Jack Morava for many precious
teachings.

Finally I would like to thank strongly Andrei Khrennikov and the
whole team at the International Center of Mathematical Modelling
in Physics and Cognitive Sciences of V\"{a}xj\"{o} for their very
generous informatics' support.

Of course nobody among the mentioned people has responsibilities
as to any (eventual) error contained in these pages.

\end{document}